\documentclass[a4paper,11pt]{article}
\pdfoutput=1

\usepackage[noblocks]{authblk}
\usepackage{here}
\usepackage{subfig}
\usepackage[pdftex]{graphicx}
\usepackage{color}
\usepackage{braket}
\usepackage{amsmath, amsfonts, amssymb, latexsym, physics}
\usepackage{siunitx}
\usepackage{feynmf}
\usepackage{bm}
\usepackage{url}
\usepackage{slashed}
\usepackage{caption}
\usepackage{tabularx}

\usepackage{jcappub, hyperref, cleveref}
\usepackage[compat=1.1.0]{tikz-feynhand}
\usepackage[samesize]{cancel}
\usepackage{fancyhdr}

\newcommand{\beq}{\begin{equation}}
\newcommand{\eeq}{\end{equation}}

\newcommand{\lmk}{\left(}
\newcommand{\rmk}{\right)}
\newcommand{\lkk}{\left[}
\newcommand{\rkk}{\right]}

\definecolor{midori}{HTML}{008000}

\begin{document}

\title{\boldmath 
Assessing the Impact of Unequal Noises and Foreground Modeling on SGWB Reconstruction with LISA
} 
\author[a,b,c]{Jun'ya Kume,}
\author[a,b]{Marco Peloso,}
\author[d]{Mauro Pieroni}
\author[e,f]{and Angelo Ricciardone}
\affiliation[a]{Dipartimento di Fisica e Astronomia ``G. Galilei'', Universit\`a degli Studi di Padova, via Marzolo 8, I-35131 Padova, Italy}
\affiliation[b]{INFN, Sezione di Padova, via Marzolo 8, I-35131 Padova, Italy}
\affiliation[c]{Research Center for the Early Universe (RESCEU), Graduate School of Science, The University of Tokyo, Hongo 7-3-1
Bunkyo-ku, Tokyo 113-0033, Japan}
\affiliation[d]{CERN, Theoretical Physics Department, Esplanade des Particules 1, Geneva 1211, Switzerland}
\affiliation[e]{Dipartimento di Fisica ``Enrico Fermi'', Università di Pisa, Largo Bruno Pontecorvo 3, Pisa I-56127, Italy}
\affiliation[f]{INFN, Sezione di Pisa,
Largo Bruno Pontecorvo 3, Pisa I-56127, Italy}

\emailAdd{junya.kume@unipd.it}
\emailAdd{marco.peloso@pd.infn.it}
\emailAdd{mauro.pieroni@cern.ch}
\emailAdd{angelo.ricciardone@unipi.it}

\subheader{{\rm RESCEU-14/24, CERN-TH-2024-170}}

\abstract{
In the search for stochastic gravitational wave backgrounds (SGWB) of cosmological origin with LISA, it is crucial to account for realistic complications in the noise and astrophysical foreground modeling that may impact the signal reconstruction.
To address these challenges, we updated the \texttt{SGWBinner} code to incorporate both variable noise levels across LISA arms and more complex foreground spectral shapes. 
We extended previous studies, which assumed only two parameters for both noise and foregrounds, simulating SGWB searches with up to 12 and 8 parameters for noise and foregrounds, respectively.
To perform this more challenging analysis, we have integrated the \texttt{JAX} framework into the \texttt{SGWBinner} code, which significantly improves its computational efficiency and enables faster Bayesian likelihood sampling and more effective exploration of complex models. 
We found that whereas increased noise complexity leads to only a tens-of-percent increase in the reconstruction error, the complexity of foregrounds can degrade the constraints by up to one order-of-magnitude depending on the assumptions.
Our findings suggest that, while moderate variations in noise amplitudes have a minimal impact, poor foreground modeling (\emph{i.e.}, templates requiring many free parameters) significantly degrades the reconstruction of cosmological signals. This underlines the importance of accurate modeling and subtraction of astrophysical foregrounds to characterize possible cosmological components.
}
\maketitle
\flushbottom

\section{Introduction}
\label{sec:introduction}

The {\it Laser Interferometer Space Antenna} (LISA)~\cite{LISA:2017pwj} is a pioneering space-based gravitational wave (GW) observatory under development by the European Space Agency (ESA) in collaboration with NASA, planned to be launched in 2035. Unlike current and planed ground-based detectors~\cite{TheLIGOScientific:2014jea,Advanced-Virgo, Aso:2013eba, LIGOScientific:2016wof,Punturo:2010zz},
which are limited to higher-frequency GW signals, LISA is specifically designed to observe the milli-Hz frequency band, opening a completely new window for GW astronomy. The observatory will consist of three satellites that approximately orbit at the vertices of an equilateral triangle with sides about 2.5 million kilometers long. By monitoring the relative displacements among the three satellites, LISA will perform three correlated interferometry measurements, allowing it to detect tens of thousands of resolvable GW events. These include mergers of Stellar Origin Binary Black Holes (SOBBHs), Compact Galactic Binaries (CGBs) mostly composed of Double White Dwarfs (DWDs), Super Massive Black Holes (SMBHs), and Extreme Mass Ratio Inspirals (EMRIs)~\cite{LISA:2022yao, Colpi:2024xhw}.

Beyond resolvable sources, numerous weak and unresolvable signals will superimpose incoherently generating a stochastic GW background (SGWB)~\cite{10.1046/j.1365-8711.2001.04217.x, Farmer:2003pa, Regimbau:2011rp,LISA:2022yao,Pozzoli:2023kxy,Babak:2023lro,Staelens:2023xjn,Toubiana:2024qxc}. At least two guaranteed components will contribute to the astrophysical SGWB in the LISA band. 
Below a few milli-Hz, the dominant contribution will come from CGBs~\cite{Evans:1987qa,Bender:1997hs}. At higher frequencies, another contribution is expected from SOBBH mergers~\cite{Lehoucq:2023zlt}, which is also probed by the operating ground-based detectors~\cite{KAGRA:2021duu,KAGRA:2021kbb}.
Moreover, recent studies have explored contributions from extreme mass-ratio inspirals (EMRIs) in the 1-10 mHz frequency range~\cite{Pozzoli:2023kxy} and extragalactic double white dwarf (DWD) systems~\cite{Staelens:2023xjn}.
All these components, typically referred to as ``confusion noise'', will act as an additional noise source in the data stream, affecting the measurements of all other signals LISA will be sensitive to\footnote{As recently shown in~\cite{Braglia:2024siw}, even a loud SGWB of cosmological origin can affect the parameter estimation significantly.}. Therefore, characterizing the astrophysical SGWB is crucial to LISA data analysis.

In addition to the astrophysical GWs, LISA will potentially be sensitive to cosmological SGWBs, which might be generated by several early universe phenomena~\cite{Caprini:2018mtu,LISACosmologyWorkingGroup:2022jok}.
Commonly and actively discussed sources for the LISA band include inflation~\cite{Bartolo:2016ami, Braglia:2024kpo}, cosmological first-order phase transitions (FOPTs)~\cite{Caprini:2015zlo, Caprini:2019egz, Giese:2021dnw,Boileau:2022ter,Gowling:2022pzb,Caprini:2024hue,Hindmarsh:2024ttn}, cosmic string networks~\cite{Auclair:2019wcv, Boileau:2021gbr,Blanco-Pillado:2024aca}, and scalar induced GWs~\cite{Matarrese:1993zf,Matarrese:1997ay,Ananda:2006af,Baumann:2007zm}(see also \cite{Domenech:2021ztg} for a review and reference therein). 
The detection and characterization of the cosmological contribution would allow us to probe high-energy particle physics beyond the Standard Model and early universe cosmology. Achieving this, however, requires precise fitting of all resolved sources and reconstruction of the astrophysical contributions. One approach currently under investigation is the simultaneous fitting of overlapping transient signals, noise components, and cosmological signals, known as a ``global fit'' scheme in the LISA data analysis~\cite{Cornish:2005qw,Vallisneri:2008ye,MockLISADataChallengeTaskForce:2009wir,Littenberg:2023xpl, Strub:2024kbe, Katz:2024oqg, Deng:2025wgk}.

A natural key question is how well LISA will distinguish cosmological SGWBs from the instrumental noise and astrophysical SGWBs (often mentioned as foreground in this context) when realistic complications in the noise and astrophysical foreground modeling are present. Typically, it is customary to assume stationarity, Gaussianity~\footnote{
In principle, the real signal might violate all these assumptions. For the impact of non-stationarity and non-Gaussianity see, \emph{e.g.,}  Refs.~\cite{Adams:2013qma,Digman:2022jmp,Mentasti:2023uyi,Hindmarsh:2024ttn,Buscicchio:2024wwm,Karnesis:2024pxh,Pozzoli:2024wfe}.
}, and the perfect knowledge of the spectral shape of these components in simulating LISA data.~\footnote{It is also common to work with purely stochastic components, assuming that all the transients are successfully removed from the data by the global fit. See Ref.~\cite{Alvey:2023npw} for the application of simulation-based inference to the SGWB search performed by LISA in the presence of transient signals.}
Many analyses assume static and equal arm lengths, as well as uniform noise amplitudes at each link connecting the spacecraft, allowing the noise spectrum to be characterized by only two parameters~\cite{Karnesis:2019mph,Caprini:2019pxz,Pieroni:2020rob,Flauger:2020qyi,Boileau:2020rpg, Dimitriou:2023knw,Caprini:2024hue,Blanco-Pillado:2024aca,Braglia:2024kpo}.
Although these assumptions reduce the complexity of the problem, they are not expected to hold in practice. Therefore, it is crucial to account for realistic complications in the noise and foreground modeling to quantify their impact on reconstructing the possible cosmological signals.
The issue has recently been a topic of active discussion in the literature. 
For noise modeling, the effect of unequal arm length and noise amplitudes on the reconstruction was recently investigated in Ref.~\cite{Hartwig:2023pft} (see also Refs.~\cite{Adams:2010vc,Adams:2013qma,Wang:2022sti} for earlier studies on the effect of unequal noise levels in the different links) and that of time-varying noise amplitudes in Ref.~\cite{Alvey:2024uoc}.
Refs.~\cite{Baghi:2023qnq, Muratore:2023gxh} explored template-based signal reconstruction while maintaining a template-free approach for the instrumental noise.
Instead, a weakly parametric approach using flexible noise and astrophysical SGWB templates was proposed in Ref.~\cite{Pozzoli:2023lgz}.

In this study, we try to put forward the understanding of how realistic complications in the noise and foregrounds affect the signal reconstruction.
For this purpose, we specifically use the \texttt{SGWBinner} code~\cite{Caprini:2019pxz,Flauger:2020qyi} to simulate the LISA data stream and perform a full likelihood sampling.
We first re-investigate the effect of unequal noise amplitude on signal parameter inference using newly implemented signal templates~\cite{Braglia:2024kpo}, instead of the flat power-law used in Ref.~\cite{Hartwig:2021mzw}.
We then consider a more general parametrization for the foreground spectral shape~\cite{Karnesis:2021tsh} in order to quantify how the signal reconstruction is affected by the simultaneous determination of the foreground spectra. 
We note that these two extensions involve a larger number of parameters than the existing analyses done with the code~\cite{Caprini:2019pxz,Flauger:2020qyi,Caprini:2024hue,Blanco-Pillado:2024aca,Braglia:2024kpo}, increasing the computational cost of the sampling process. 
To mitigate such an increase in the computational cost, we have integrated the \texttt{JAX} library~\cite{jax2018github} into the existing code, which accelerated the code of a factor up to 10 with respect to its previous version. 

This paper is organized as follows. In Sec.~\ref{sec:detection}, we review the formalism for modeling the LISA data stream and introduce the noise and foreground modeling used in our updated \texttt{SGWBinner} code. We then summarize the analysis scheme of the \texttt{SGWBinner} code and briefly describe our update.
In Sec.~\ref{sec:extended_models}, we report the result of new analyses performed with the accelerated code. 
The effect of unequal noise level and unfixed foreground shape on the signal parameter inference is discussed in Sec.~\ref{sec:uneq_noise} and Sec.~\ref{sec:extend_fg}, respectively.
Finally, Sec.~\ref{sec:discussion} is devoted to the discussion of these results.

\section{LISA data modeling and analysis in \texttt{SGWBinner}}
\label{sec:detection}
In this section, we illustrate our model for the data LISA will collect. After briefly reviewing some aspects of the SGWB search with LISA, we discuss the noise and foreground model adopted in our analysis. 
Then we present the key ingredients of the \texttt{SGWBinner} code~\cite{Caprini:2019pxz,Flauger:2020qyi} and its analysis routine, whose implementation closely follows the description and notation provided in this section.

\subsection{SGWB search with LISA TDI channels}
\label{sec:power_spectra}

LISA will monitor the fractional Doppler frequency shifts induced by GWs on photons traveling along the arms of the detector. In this context, the one-way path connecting two satellites is typically dubbed ``link''. 
To suppress laser frequency noise, which is expected to be several orders of magnitude greater than the required sensitivity~\cite{LISA:2017pwj}, LISA will employ Time-Delay Interferometry (TDI)~\cite{Armstrong_1999,Prince:2002hp,Shaddock:2003bc,Shaddock:2003dj,Tinto:2003vj,Vallisneri:2005ji,Muratore:2020mdf,Tinto:2020fcc,Muratore:2021uqj}. TDI is a post-processing technique to combine the six link measurements into data channels where laser frequency noise is strongly suppressed. 
Since, for simplicity, we assume equal and static arm lengths, the so-called first-generation TDI variables suffice to achieve laser noise cancellation~\footnote{For the realistic orbit where arm lengths vary over time, one can utilize the second-generation TDI variables~\cite{Vallisneri:2005ji,Muratore:2020mdf,Muratore:2021uqj,Hartwig:2021mzw}.
While we did not test it explicitly, we expect that the main conclusion --that foreground modeling is more critical than noise modeling-- would remain
valid for the second-generation variables under the same hypothesis. In practice, higher-generation TDIs introduce more loops in the optical path. Since this operation affects the signal and the noise parts similarly, \emph{i.e.}, introducing additional zeros in the response functions, we do not expect any significant difference in the results derived in this work.}. 

Denoting the single link measurement as $\eta_{\alpha\beta}(t)$, where $\alpha\beta \in \left\{12,23,31,21,32,13\right\}$ representing the pairs of satellites $\alpha \beta$ (with the laser emitted from the satellite $\beta$ at time $t - L_{\alpha \beta}/c$ and recorded at time $t$ in satellite $\alpha$, and $L_{\alpha \beta}$ the unperturbed geodesic distance between the two satellites),
the commonly used Michelson TDI combinations $\left\{\rm X,Y,Z\right\}$ are expressed as 
\begin{equation}
    {\rm X} \equiv (1 - D_{13}D_{31})(\eta_{12} + D_{12}\eta_{21}) + (D_{12}D_{21} - 1)(\eta_{13} + D_{13}\eta_{31})\,,
\end{equation}
where ${\rm Y}$ and ${\rm Z}$ are obtained by cyclic permutations of the indices for the satellite pairs in the definition of ${\rm X}$.
Here $D_{\alpha\beta}$ is the delay operator acting on any time-dependent function $x(t)$ as $D_{\alpha\beta} \, x(t) = x(t - L_{\alpha\beta})$. Notice that in practice, TDI can be understood as the operation of $3\times6$ matrix on the six single link measurements that returns the three TDI channels in a given basis, regardless of the generation as explicitly presented in Refs.~\cite{Baghi:2023qnq,Hartwig:2023pft}.
In the following, we assume equal arms, {\it i.e.}, we take $L_{\alpha\beta}=L=2.5 \times 10^9\,{\rm m}$. 

It is convenient to combine the TDI variables to obtain the so-called AET basis~\cite{Hogan:2001jn,Adams:2010vc}, defined as
\begin{equation}
\label{eq:XYZ_to_AET}
    \rm{A} \equiv \frac{{\rm Z-X}}{\sqrt{2}}, \qquad
    \rm{E} \equiv \frac{{\rm X - 2Y + Z}}{\sqrt{6}}, \qquad
    \rm{T} \equiv \frac{{\rm X + Y + Z}}{\sqrt{3}} \; , 
\end{equation}
which, in the limit of equal arms and equal noises, can be shown to be orthogonal ({\it i.e.}, to have vanishing cross-correlations). Moreover, as originally noticed in~\cite{Prince:2002hp}, due to its symmetric structure, the T channel strongly suppresses GW signals at small frequencies, where it is effectively noise-dominated (see, \emph{e.g.}, Fig.~3 of~\cite{Flauger:2020qyi}, for a quantitative comparison). 
For this reason, the T channel is typically referred to as a null channel, which can be used for noise monitoring.
While these properties, make the AET basis particularly convenient for SGWB searches,  
for the moment let us proceed with an arbitrary basis to keep the generality of our discussion.

We denote with $d_i(t)$ the three time-domain data streams, where $i$ runs over the channels of the TDI basis. For notational convenience, we model it as a real-valued function on the interval $\left[ -\tau/2, \tau/2\right]$ with $\tau$ being the duration of a data segment.
Then, the Fourier transforms of the data streams are
\begin{equation}
\tilde{d}_i\left(f\right) =  \int_{-\tau/2}^{
\tau/2} {\rm d} t  \;\textrm{e}^{ 2\pi i f t} d_i \left(t\right)\,.
\end{equation}
Throughout this paper (and also in the code), the data is assumed to be `perfect' residuals.
That is, all loud deterministic signals and glitches in the noise are assumed to be subtracted from the time stream. This could be achieved through some appropriate methods within the LISA global fit scheme~\cite{Cornish:2005qw,Vallisneri:2008ye,MockLISADataChallengeTaskForce:2009wir,Littenberg:2023xpl, Strub:2024kbe, Katz:2024oqg} and the glitch subtraction methods~\cite{Robson:2018jly,Baghi:2021tfd,Spadaro:2023muy,Houba:2024tyn}.
After this procedure, the data only contains the noise $\tilde{n}_i^{\nu}$ and the residual stochastic signals $\tilde{s}_i^{\sigma}$ 
\begin{equation}
    \tilde{d}_i(f) = \sum_{\nu}\tilde{n}_i^{\nu}(f) + \sum_{\sigma}\tilde{s}_i^{\sigma}(f),
\end{equation}
where $\nu, \sigma$ run over the different noise and signal components, respectively.
Assuming that all these components obey stationary and Gaussian statistics, the ensemble average of the Fourier modes is characterized by
\begin{equation}
\langle \tilde{n}_i^{\nu}(f) \tilde{n}_j^{\nu*}(f') \rangle = \frac{1}{2} \delta \left(f-f'\right) P^{\nu}_{N,ij}(f) \;, \quad\quad 
\langle \tilde{s}_i^{\sigma}(f) \tilde{s}_j^{\sigma*}(f') \rangle = \frac{1}{2} \delta \left(f-f'\right) P^{\sigma}_{S,ij}(f) \;,
\end{equation}
where we define the one-side power-spectral density (PSD) (for $i = j$) and cross-spectral density (CSD) (for $i \neq j$) of noise and signal components $P^{\nu}_{N,ij}(f)$ and $P^{\sigma}_{S,ij}(f)$, respectively. 
Note that by definition, these are Hermitian matrices with respect to the indices $ij$. 
Assuming all these components to be uncorrelated with one another, we obtain
\begin{equation}
\begin{aligned}
\label{eq:singlesidedsps}
\langle \tilde{d}_i(f) \tilde{d}_j^*(f') \rangle &= \frac{1}{2} \delta \left(f-f'\right) \lkk \sum_{\nu}P^{\nu}_{N,ij}(f) +\sum_{\sigma}P^{\sigma}_{S,ij}(f)\rkk \\
&\equiv \frac{1}{2} \delta \left(f-f'\right)
\lkk P_{N,ij}(f) + P_{S,ij}(f)\rkk \; ,
\end{aligned}
\end{equation}
where $P_{N,ij}(f)$, $P_{S,ij}(f)$ are the total noise and signal PSDs and CSDs.

At this point, let us introduce the response functions for isotropic SGWB signals $\mathcal{R}_{ij}(f)$~\cite{Flauger:2020qyi}.
Under the assumption of equal and static arm, the response functions can be expressed as
\begin{equation}
\mathcal{R}_{ij}(f) = 16 \sin^2\left(\frac{f}{f_c}\right) \left(\frac{f}{f_c} \right)^{2} \tilde{R}_{ij}(f) \; , \qquad \label{eq:response_geom}
\end{equation}
where we have introduced the detector characteristic frequency $f_c \equiv (2\pi L/c)^{-1} \simeq 19$mHz and again $i,j$ are TDI indexes. 
The last factor in this equation, {\it i.e.}, the $\tilde{R}_{ij}(f)$, encodes the projection of the tensorial structure onto the geometry of the detector. Approximated expressions for this quantity for the XYZ and AET TDI basis are reported, {\it e.g.}, in~\cite{Robson:2018ifk,Flauger:2020qyi}, where one finds that, under the assumptions stated above in this section, $\tilde{R}_{ij}(f)$ is diagonal in the AET basis. 
The responses project the SGWB (in either strain $S^{\sigma}_h(f)$ or Omega units $\Omega_{\rm GW}^{\sigma}(f)$) onto the data PSDs and CSDs as
\begin{equation}
		\label{eq:signal_psd}
		P_{S,ij}(f) =  \mathcal{R}_{ij} (f) \sum_{\sigma}S_h^{\sigma} (f) \; = \mathcal{R}_{ij} (f) \frac{3H_0^2}{4\pi^2f^3}\sum_{\sigma}h^2\Omega_{\rm GW}^{\sigma}(f),
\end{equation}
where $H_0$ is the present Hubble constant and $h$ is the normalized one as $H_0/h \simeq 3.24 \times 10^{-18}$ 1/s.
It is common practice to predict the primordial SGWB signal in terms of $h^2 \Omega_{\rm GW}(f)$; therefore, for later convenience, we define
\begin{equation}
\label{eq:omega_sens}
		P_{N,ij}^{\Omega}(f) = \frac{4 \pi^2 f^3}{3 H_0^2} P_{N,ij}(f) \;.
\end{equation}
In the following, we will give more detailed descriptions of the noise sources in the two TDI bases $\left\{\rm X,Y,Z\right\}$ and $\left\{\rm A, E, T\right\}$,
and of the astrophysical foregrounds which are included in $h^2\Omega_{\rm GW}^{\sigma}(f)$.

\subsubsection{LISA noise model}
\label{sec:noise_and_signal_model}
As discussed in the previous section, the TDI variables are designed to eliminate the dominant laser frequency noise. 
In a simplified approach, the residual noise components (dubbed secondary noise) that enter into each TDI channel can be grouped into two effective quantities, namely, ``Optical Metrology System" (OMS) noise and ``Test Mass" (TM) noise.
The former accounts for noise in the readout frequency, due, for example, to laser shot noise, while the latter is associated with the random displacements of the test masses caused, for example, by local environmental disturbances.
Introducing the transfer functions for these two noise sources $\mathcal{T}^{\nu}_{ij,\alpha\beta}(f)$ (for details, see, {\it e.g.},~\cite{Flauger:2020qyi,Hartwig:2021mzw,QuangNam:2022gjz,Hartwig:2023pft}), which project those contributions onto the TDI channels,
the total noise power spectrum can be expressed as 
\begin{equation}
		\label{eq:two_signal_correlators}
		P_{N,ij}(f) = \sum_\nu P_{N,ij}^{\nu}(f)
  = \sum_{\alpha\beta}\lkk \mathcal{T}_{ij,\alpha\beta}^{\rm TM} (f) S_{\alpha\beta}^{\rm TM} (f) 
  + \mathcal{T}_{ij,\alpha\beta}^{\rm OMS} (f) S_{\alpha\beta}^{\rm OMS} (f)
  \rkk .
\end{equation}
Our current knowledge of the LISA noise is based on the LISA Pathfinder~\cite{PhysRevLett.116.231101} and laboratory tests. Since the precise determination of noise properties is one of the main technical challenges of the future LISA mission and lays beyond the scope of the current work, 
as customary in the literature, we assume stationary, Gaussian, and uncorrelated noises at each link with identical spectral shapes given by~\cite{ldcdoc}
\begin{align}
\label{eq:TM_noise_def}
S^\text{TM}_{\alpha\beta}(f) & = 7.737\times 10^{-46} \times A_{\alpha\beta}^2 \lmk \frac{f_c}{f} \rmk^2 \;  \left[1 + \left(\frac{0.4 \textrm{mHz}}{f}\right)^2\right]\left[1 + \left(\frac{f}{8  \textrm{mHz}}\right)^4\right] \;  \times \mathrm{s} \;,  \\ 
\label{eq:OMS_noise_def}
S^\text{OMS}_{\alpha\beta}(f) & = 1.6 \times 10^{-43} \times P_{\alpha\beta}^2 \lmk \frac{f}{f_c} \rmk^2 \; \left[1 + \left(\frac{2\textrm{mHz}}{f}\right)^4 \right] 
\;  \times \mathrm{s}  \;,
\end{align}
where $A_{\alpha\beta}$ and $P_{\alpha\beta}$ are parameters characterizing the amplitudes of the TM and OMS noises in the different links. In App.~\ref{sec:full_noise}, we present the full expressions of $P_{N,ij}$ for the first-generation TDI variables in the XYZ and AET bases.

For simplicity, most studies in the literature assume the noise amplitudes to be identical, {\it i.e.}, $A_{\alpha\beta} = A$ and $P_{\alpha\beta} = P$, and according to the ESA mission specifications the face values are $A= 3$ and $P=15$~\cite{Colpi:2024xhw}. 
In this case, noise spectra reduce to $S^\text{TM}_{\alpha\beta}(f) = S^\text{TM}(f, A)$ and $S^\text{OMS}_{\alpha\beta}(f) = S^\text{OMS}(f, P)$ and in the AET basis, $P_{N,ij}(f, A, P)$ becomes diagonal, with identical AA and EE components (see Eq.~\eqref{eq:psdAA}). As the signal response also becomes diagonal, the numerical evaluation of the likelihood function is simplified in this basis (see Sec.~\ref{sec:data_analysis}).
However, the equal noise and equal arm length assumptions are quite idealized. 
As for the former, one can naturally imagine the presence of small differences among the noises in the six links. Regarding the latter, it is known that, with realistic orbits, LISA will not be perfectly equilateral with non-static arm-lengths that vary at the percent level~\cite{Martens:2021phh} (see also Appendix A of~\cite{Mentasti:2023uyi}).

To simulate the LISA SGWB search with a more realistic noise property, the effect of unequal noise amplitude and unequal arm length on the signal parameter inference has been recently studied in Ref.~\cite{Hartwig:2023pft}. For a flat power-law signal, it has been shown that the amplitude reconstruction is almost unaffected.
As reported in Sec~\ref{sec:uneq_noise}, we performed analyses under the same assumptions of different noise amplitudes in the different spacecrafts and tested them for different cosmological signals.

\subsubsection{Astrophysical foregrounds}\label{sec:foreground}
As already mentioned in Sec.~\ref{sec:introduction}, apart from possible cosmological sources, there are at least two guaranteed components contributing to the astrophysical SGWB signal in the LISA band, {\it i.e.}, the SGWB from CGBs and SOBBHs.
Consequently, in the following, we assume the total signal power can be expressed as
\begin{equation}
	\label{eq:signal_psd_components}
	h^2 \Omega_{\rm GW}(f) = 
    h^2\Omega^{\rm Gal}_{\rm GW}(f) + h^2\Omega^{\rm Ext}_{\rm GW}(f) + 
    h^2\Omega^{\rm Cosmo}_{\rm GW}(f),
\end{equation}
where the first two components represent those astrophysical ``foregrounds'' and the last one represents the contribution from cosmological sources. 
We leave the inclusion of other possible contributions ({\it e.g.}, from EMRIs~\cite{Pozzoli:2023kxy} and DWDs~\cite{Staelens:2023xjn}) in our code to future work. 
In the remainder of this section, we provide the templates for these foreground components implemented in the \texttt{SGWBinner} code that have recently been used in Refs.~\cite{Caprini:2024hue,Blanco-Pillado:2024aca,Braglia:2024kpo}. 

\subsubsection*{Galactic foreground}
After the removal of loud signals produced from the population of CGBs in the galactic disk~\cite{Nissanke:2012eh}, there remains a strong stochastic component, dubbed Galactic foreground, consisting of the unresolved sub-threshold mergers of CGBs (mostly, DWDs). 
Due to the anisotropic distribution of the unresolved sources, the angular dependence of the response functions and the yearly orbit of LISA, this component is known to have an annual modulation.
In principle, this feature can be used to separate the Galactic component from the other contributions considered to be stationary, {\it e.g.}, by properly taking into account the variation in each chunk~\cite{Adams:2013qma,Mentasti:2023uyi,Hindmarsh:2024ttn}, or by accounting the frequency correlation arising from the (parameterized) annual modulation~\cite{Pozzoli:2024wfe}.
However, reconstruction with such a procedure is inevitably sensitive to uncertainties due to the non-stationarity of noise and data gaps.
Therefore, instead of adopting such a strategy, we only consider the signal integrated over the whole sky and observation time, modeling and treating it as an isotropic component.
As the temporal effect of those uncertainties is expected to be mitigated with the average over several years, this approach can be regarded as reasonable yet conservative\footnote{By averaging over the observation time and angles, the information on the angular structure is lost. As a consequence, we rely only on the frequency spectra to separate the different components. Including that information would lead to better results, but at a higher computational cost. For this reason, we consider the results of our analysis to be conservative.}.

This integrated contribution can be described by the following empirical model studied in Ref.~\cite{Karnesis:2021tsh}:
\begin{equation}\label{eq:gal}
    h^2\Omega^{\textrm{Gal}}_{\rm GW}(f)=\frac{1}{2}\left(\frac{f}{ 1\,\textrm{Hz}}\right)^{n_{\rm Gal}}
    e^{-(f/f_1)^\alpha}
    \left[1+\tanh{\frac{f_{\textrm{knee}}-f}{f_2}}\right]h^2\Omega_{\textrm{Gal}}\;,
\end{equation}
where the value of $f_1$ and $f_{\textrm{knee}}$ depends on the effective observation time $T_{\rm Obs}$ as
\begin{equation}
\begin{aligned}
    \log_{10} (f_1/{\rm Hz}) &= a_1 \log_{10}(T_{\rm Obs}/{\rm year}) + b_1\,,\\
    \log_{10} (f_{\textrm{knee}}/{\rm Hz}) &= a_k \log_{10}(T_{\rm Obs}/{\rm year}) + b_k\,.
\end{aligned}
\end{equation}
The exponential factor $e^{-(f/f_1)^{\alpha}}$ is due to the loss of stochasticity at higher frequency where less signals are superimposed at the same time~\cite{Karnesis:2021tsh}.
The last $\tanh$ term is modeling the expected complete subtraction of CGBs signal at frequencies $f > f_{\rm knee}$.
Based on Ref.~\cite{Karnesis:2021tsh}, we adopt the fiducial values $a_1 = -0.15$, $b_1 = -2.72$, $a_k = -0.37$, $b_k = -2.49$, $\alpha = 1.56$, $f_2 = 6.7\times10^{-4}$Hz, $\log_{10}(h^2\Omega_{\rm Gal}) = -7.84$ and $n_{\rm Gal} = 2/3$, which is expected from the superposition of inspiraling binaries. 

In the previous analyses reported in Refs.~\cite{Caprini:2024hue,Blanco-Pillado:2024aca,Braglia:2024kpo}, only the amplitude parameter $h^2\Omega_{\rm Gal}$ was fitted in the signal reconstruction, assuming a complete knowledge of its spectral shape. Given the uncertainty of the above empirical model, we relax such an assumption and, in our analysis, in addition to $h^2\Omega_{\rm Gal}$, we also vary $n_{\rm Gal}$, $f_1$, $f_2$, $f_{\rm knee}$ and $\alpha$. The effect of fitting these parameters on the signal reconstruction is discussed in Sec.~\ref{sec:extend_fg}.

\subsubsection*{Extra-galactic foreground}
The other astrophysical component in eq.~(\ref{eq:signal_psd_components}), referred to as Extra-galactic foreground, is the incoherent superposition of all the extra-galactic compact object mergers. The contributions from SOBBHs and binary neutron stars (BNS) in their inspiral phase, which cannot be individually resolved by LISA~\cite{Sesana:2016ljz}, are estimated through the observation of ground-based detectors~\cite{KAGRA:2021duu,KAGRA:2021kbb}. 
Due to the isotropic distribution of the sources and the limited angular resolution of LISA, this foreground can be well modeled as an isotropic SGWB signal with the power-law shape~\cite{Sesana:2016ljz, LIGOScientific:2019vic}: 
\begin{equation}\label{eq:ext}
        h^2 \Omega^{\rm Ext}_{\rm GW}(f)  = h^2\Omega_{\rm Ext} \left( \frac{f}{1 {\rm mHz}}\right)^{n_{\rm Ext}} \; ,
\end{equation}
where $h^2 \Omega_{\rm Ext}$ and $n_{\rm Ext}$ are the amplitude (at 1mHz) and tilt of the spectrum. From the recent observation by LIGO-Virgo-KAGRA collaboration, the magnitude of the SGWB signal from SOBBHs and BNS is estimated as $\Omega_{\rm Ext} =  7.2^{+3.3}_{-2.3}  \times 10^{-10} $ at $f = 25$\,Hz~\cite{KAGRA:2021duu,Babak:2023lro}. 
Extrapolating this amplitude in the LISA frequency band\footnote{
Several subtleties will have to be addressed to make this model more accurate. It is for instance not obvious that differences in the performance of the LISA and LIGO–Virgo–KAGRA detectors to resolve signals from SOBBHs play no relevant role, or that the binary eccentricity evolution is a minor effect (see, {\it e.g.}, Refs.~\cite{DOrazio:2018jnv, Zhao:2020iew}).}, we adopt the fiducial value $\log_{10}(h^2\Omega_{\rm Ext}) = -12.38$ and $n_{\rm Ext} = 2/3$.

Once again, previous studies only considered $h^2\Omega_{\rm Ext}$ as a free parameter to be fitted together with cosmological parameters. Similarly to the galactic component, we also vary $n_{\rm Ext}$ in our analysis and discuss its effect in Sec.~\ref{sec:extend_fg}.\\

\subsection{Mock data analysis with the \texttt{SGWBinner} code}\label{sec:data_analysis}
In this section, we summarize the data analysis scheme implemented in the \texttt{SGWBinner} code, for more details see~\cite{Flauger:2020qyi}. Let us start with the data generation. As described in Refs.~\cite{Caprini:2019pxz,Flauger:2020qyi}, given the total observation time $T_{\rm Tot}$ and duty cycle $\mathcal{D}_c$ of the experiment the code computes the effective observation time $T_{\rm Obs} = \mathcal{D}_c \times T_{\rm Tot}$ and generates data in the frequency domain for a given number of segments $N_d$ with duration $\tau = T_{\rm Obs} / N_d$. 
That is, given the stationarity of our model, $N_d$ Gaussian realizations for the signal, noise, and foregrounds are generated independently at each frequency bin $f_k$ (the frequency band is assumed to cover the whole LISA band $[3\times10^{-5},0.5]$Hz with spacing $\Delta f = 1/\tau$), with zero mean and variances defined by their respective power spectral densities.
Starting from the data $\tilde{d}^s_i(f_k)$, where $s$ indexes segments, we compute the estimated PSD/CSD as $\bar{D}_{ij}^k \equiv \sum_{s=1}^{N_d} d_i^s(f_k)d_j^{s*}(f_k)/N_d$ by averaging over time segments.
To lower the numerical complexity, the code then performs coarse-graining over the frequency with inverse variance weighting, producing a data set $D_{ij}^k$ where $k$ runs over a smaller set of frequency bins $f_{ij}^k$ (namely, the coarse-graining reduces the number of bins, increasing their size). Notice that $D_{ij}^k$ retains statistical properties similar to those of $\bar{D}_{ij}^k$. Similarly to Refs.~\cite{Flauger:2020qyi,Caprini:2024hue,Blanco-Pillado:2024aca,Braglia:2024kpo}, we set $\tau = 11.4$ days (corresponding to $\Delta f = 10^{-6}$Hz) and $N_d = 126$ in our analysis, implying $T_{\rm Obs} = 4$ effective years of data.
Since LISA is scheduled to function for at least 4.5 years up to a maximum of 10 years (4.5 years $\lesssim T_{\rm Tot} \lesssim$ 10 years) and duty cycle will be at most $\mathcal{D}_c =  82$\%~\cite{Colpi:2024xhw}, our choice $T_{\rm Obs} = 4$ is relatively conservative.

After generating simulated data employing the noise models described above (as well as the signal templates that we discuss below) the code attempts to reconstruct signals and estimate their errors by working on the posterior distribution for the model parameters defined as
\begin{equation}
	p(\vec{\theta}|D) \equiv \frac{\pi(\vec{\theta})  \mathcal{L}(D | \vec{\theta})} {\mathrm{Z}(D) }\; ,\label{eq:full_post}
\end{equation}
where $\mathcal{L}(D|\vec{\theta}) $ is the likelihood of the data $D$, $\pi(\theta)$ is the prior for the parameters $\vec{\theta}$, and $\mathrm{Z} (D) $ is the model evidence. 
Here the vector of parameters $\vec{\theta}$ is expressed as
\begin{equation}
    \vec{\theta} \equiv \{\vec{\theta}_{\rm cosmo}, \vec{\theta}_n, \vec{\theta}_{\rm fg} \},
\end{equation}
where the components represent signal, noise, and foreground parameters, respectively. In practice, in Refs.~\cite{Flauger:2020qyi,Caprini:2024hue,Blanco-Pillado:2024aca,Braglia:2024kpo}, the cosmological parameters were assumed to be template-dependent (see App.~\ref{sec:templates}), and only two noise ({\it i.e.}, $\vec{\theta}_n = \{A, P\} $) and two foreground parameters ({\it i.e.},  $\vec{\theta}_{\rm fg} = \{ \log_{10}(h^2\Omega_{\rm Gal}), \log_{10}(h^2\Omega_{\rm Ext}) \} $) were considered. 
As anticipated, we will go beyond the latter assumptions. 

The likelihood employed in the code reads~\cite{Flauger:2020qyi,Caprini:2024hue,Blanco-Pillado:2024aca,Braglia:2024kpo}
\begin{equation}
\ln  \mathcal{L}(D|\vec{\theta}) = \frac{1}{3} \ln \mathcal{L}_{G}(D|\vec{\theta})+ \frac{2}{3} \ln  \mathcal{L}_{LN}(D|\vec{\theta}) \; ,\label{eq:full_likelihood}
\end{equation}
with
\begin{equation}
    \label{eq:gaussian_likelihood}
	\ln \mathcal{L}_G(D|\vec{\theta}) = - \frac{ N_d }{2} \sum_{i,j} \sum_{k} n_{ij}^{(k)} \left[ \frac{ \mathcal{D}_{ij}^{th}(f_{ij}^{(k)}, \vec{\theta}) -\mathcal{D}_{ij}^{(k)}}{ \mathcal{D}_{ij}^{th}(f_{ij}^{(k)}, \vec{\theta})} \right]^2 \;,
\end{equation}
\begin{equation}
\label{eq:lognormal_likelihood}
\ln \mathcal{L}_{LN}(D|\vec{\theta}) = - \frac{N_d}{2} \sum_{i,j} \sum_{k} n_{ij}^{(k)}  
\ln ^2 \left[ \frac{ \mathcal{D}_{ij}^{th}(f_{ij}^{(k)}, \vec{\theta})  }{ \mathcal{D}_{ij}^{(k)} } \right]  \; , 
\end{equation}
where the indices $i,j$ run over the TDI channels, the index $k$ runs over the coarse-grained data points, and $n_{ij}^{(k)}$ represents the number of points within bin $k$ for the cross-spectrum of channels $i$ and $j$.
Here the theoretical model for the data including all SGWB and noise components reads $ \mathcal{D}_{ij}^{th}(f, \vec{\theta}, \vec{n}) \equiv  \mathcal{R}_{ij} \, 
h^2\Omega_{\rm GW}(f, \vec{\theta}_{\rm cosmo}, \vec{\theta}_{\rm fg}) +  P_{N,ij}^{\Omega}(f, \vec{\theta}_{\rm n})$. 
Assuming the equal and static arm length and equal noise, the previous studies worked in the AET basis as $\mathcal{D}_{ij}$ become diagonal.
Notice that when these assumptions are violated, which will be the case with real data, the analysis only involving the diagonal parts becomes suboptimal.
This issue will be addressed in Sec.~\ref{sec:uneq_noise}.

In practice, the \texttt{SGWBinner} code allows both template-free (or bin-by-bin) reconstruction~\cite{Caprini:2019pxz,Flauger:2020qyi} and template-based reconstruction~\cite{Caprini:2024hue,Blanco-Pillado:2024aca,Braglia:2024kpo}.
While the former agnostic search is flexible, it is (obviously) suboptimal for reconstructing any given template. 
The present study is concerned only with template-based signal reconstruction, also bearing in mind the possibility that a previous agnostic search has provided support for a specific frequency shape of the signal, that has the potential to be well described by the template under consideration. While the code supports both Fisher analysis and Bayesian inference with likelihood sampling, in this work, we only perform the latter. 
For this purpose, we used the nested sampler \texttt{Polychord}~\cite{Handley_2015,Handley:2015fda}, via its  \texttt{Cobaya}  interface and visualized the results using \texttt{GetDist}~\cite{Lewis:2019xzd}. Note that when fitting the simulated data, the same noise and foreground model used to generate the data is applied. This is particularly crucial for LISA, where 
noise and foreground components must be fitted simultaneously with the reconstruction of the signals.
This means that any discrepancies between the actual instrumental/foreground noise and their models used in the inference could introduce bias. 
This issue will have to be closely monitored in future upgrades of the code.

\subsection{Integration of \texttt{JAX} framework}
When the number of parameters is large, the sampling can be computationally very expensive. This is the case for our study where we loosen the assumptions made on the noise and foregrounds. Namely, we fit 12 parameters for the noise instead of 2 as reported in Sec.~\ref{sec:uneq_noise} and 6 additional parameters for the foregrounds as in Sec.~\ref{sec:extend_fg}, respectively. 
In order to obtain better predictions within a reasonable computation time, we worked on the acceleration of the sampling process in the \texttt{SGWBinner} code, 
by interfacing the \texttt{JAX} framework.
Here we briefly describe how the \texttt{JAX} framework and the JIT compilation can accelerate a Python code and how to accommodate it into the \texttt{SGWBinner} code.

In spite of the flexibility in the coding, Python is affected by a slower execution time as compared to compiled languages such as C/C++. To overcome this issue, the \texttt{JAX} library utilizes Just-In-Time (JIT) compilation that traces the execution of a given Python function the first time it is called and compiles it into a faster executable. Such a process is highly beneficial, for example, in the MCMC sampling where the same likelihood function is called and computed over an extremely large number of times.
This process is handled with the XLA compiler which is highly optimized for CPU, GPU, and TPU execution. Specifically for \texttt{JAX} which includes bespoke re-implemented packages such as \texttt{jax.numpy} and \texttt{jax.scipy}, existing Python codes, {\it e.g.}, based on \texttt{Numpy} and/or \texttt{SciPy}, can easily be converted to be compatible with the XLA compiler.
Therefore, we prepared a new version of the existing template bank and the library for noise rewritten from \texttt{NumPy} to \texttt{jax.numpy}, from which the JIT-compatible likelihood instance is constructed. 
With this update, the \texttt{SGWBinner} code is now able to perform the JIT-compilable computation of the likelihood in Eq.~\eqref{eq:full_likelihood}.

The updated code performs Bayesian inference via sampling by passing this JIT-compilable likelihood to the \texttt{Cobaya} interface. 
As a result, the overall computation time for the sampling process is reduced in direct proportion to the improvement in the likelihood evaluation speed.
As summarized in App.~\ref{sec:accelerate}, depending on the cosmological signal template, the evaluation is now accelerated up to 10 times, highly benefiting the analyses that we describe below.
We note that all computations in our analyses are performed on the CPU for compressed data. The potential utilization of more advanced features of the \texttt{JAX} library, including GPU acceleration, is discussed in Sec.~\ref{sec:discussion}.

\section{Assessing the impact of unequal noise and foreground modeling}\label{sec:extended_models}
In this section, we present the result of our new analyses based on the more complex noise and foreground models discussed in Sec.~\ref{sec:power_spectra}.
In the examples below, we adopted the log-normal bump model (see Eq.~\eqref{eq:log-normal_bump}) as a proxy for the cosmological signal. 
The reason for the choice of this template is twofold. Firstly, it describes an increase in the SGWB power only over a limited range of frequencies, as it might have been caused by a number of cosmological mechanisms~\cite{Braglia:2024kpo} active only at those frequencies (for instance, a sudden episode of particle production during inflation, leaving a marked signal only at the scales that left the horizon at that moment). Therefore, analyses with this template can provide generic suggestions for scenarios that source GWs in a finite time and predict a peaked spectrum. Secondly, despite its simplicity, this template suffices to highlight the impact of unequal noise amplitudes and more elaborate foregrounds in measuring the template parameters. 

Following Ref.~\cite{Braglia:2024kpo}, we hereafter assume the log-uniform priors for those signal parameters as $h^2\Omega_* \in [10^{-30}, 10^{-5}], \rho \in [10^{-2},10], f_* \in [10^{-5},10^{-1}]$Hz.
On the other hand, similarly to Refs.~\cite{Flauger:2020qyi,Caprini:2024hue,Blanco-Pillado:2024aca,Braglia:2024kpo}, we assume Gaussian priors for noise amplitudes (both in the unequal case of Sec.~\ref{sec:uneq_noise} and in the equal case of Sec.~\ref{sec:extend_fg}) and the foreground amplitudes. For the priors of noise amplitudes, we set the means to be the fiducial values and the standard deviations to be 20\% of the fiducial values. 
For the foreground amplitudes $\log_{10}h^2\Omega_{\rm Gal}$ and $\log_{10}h^2\Omega_{\rm Ext}$, consistently with previous studies in the literature~\cite{Caprini:2024hue, Blanco-Pillado:2024aca, Braglia:2024kpo}, we set the means to be the fiducial values and the standard deviations to be 0.21 and 0.17, respectively. While for the extragalactic component, this number is motivated by the estimates performed in Ref.~\cite{Babak:2023lro}, for the galactic component, given that the statistical uncertainty is typically of the order of $10^{-2}$ (see, {\it e.g.}, Fig.~\ref{fig:uneq_1D}), this effectively corresponds to choosing an informative prior. 

Our main results, {\it i.e.}, the corner plots are supplemented with the plots of power spectral density of injected/reconstructed noise, signal and foregrounds in the Omega units. 
As a simple indicator of the reach of LISA, we also compute the signal-to-noise ratio (SNR) defined as
\begin{equation}
\label{eq:SNR_def}
\textrm{SNR} \equiv \sqrt{T_{\rm Obs} \;\sum_{i \in \{ {\rm AET} \} } \int_{f_{\rm min}}^{f_{\rm max}} \textrm{d}f \; \left( \frac{P^{\sigma}_{S,ii}}{P_{N,ii}} \right)^2}  \; ,
\end{equation}
and plot the power-law sensitivity (PLS)~\cite{Thrane:2013oya, Caprini:2019pxz} for SNR = 10 and $T_{\rm Obs} = 4$ years. Notice that the PLS only serves as a graphical indicator suggesting that the signal can be sufficiently loud to be seen by the detector. All our statements concerning the detectability of signals and statistical uncertainties are based on data analysis techniques.

\subsection{Unequal noise amplitudes}\label{sec:uneq_noise}
For this analysis, we implemented the general expressions of noise spectrum reported in Eqs.~\eqref{eq:TM_noise_def}--~\eqref{eq:OMS_noise_def} into the \texttt{SGWBinner} code. 
Notice that in this generalized scenario, the AET basis remains diagonal in the signal but not in the noise part (see App.~\ref{sec:full_noise} for the non-vanishing elements of the noise cross-spectrum). As a consequence, excluding the off-diagonal terms, which contain additional information on the noise,  results in a sub-optimal estimation of the noise parameters.  Since there are non-zero correlations between the noise and signal parameters, this approach will also degrade the precision with which the signal parameters are determined.  In this sense, our analysis, which is carried out in the AET basis without including the off-diagonal terms is suboptimal. 
In this work, for simplicity, we restrict ourselves to this suboptimal analysis, leading to results that can be seen as conservative. 
Similarly to Ref.~\cite{Hartwig:2023pft}, we set the fiducial values of the noise parameters $A_{\alpha\beta}$ and $P_{\alpha\beta}$ to
\begin{equation}
\begin{aligned}
A_{\alpha\beta}  &=  \{3.61, 3.02, 2.87, 3.43, 2.65, 3.45 \},\\ 
P_{\alpha\beta}  &=  \{ 14.00, 16.93, 9.43, 21.55, 17.04, 20.83 \},
\end{aligned}
\label{eq:unequal_noise_levels}
\end{equation}
which were drawn from Gaussian distributions centered on the fiducial values $A_0 = 3$ and $P_0 = 15$, respectively, with standard deviations equal to $20\%$ of the fiducial values.
From the values in Eq.~\eqref{eq:unequal_noise_levels}, the compressed data was generated as discussed in Sec.~\ref{sec:data_analysis}. 
In doing so, we assume time-independent statistical properties of the noise. 
At the end of this subsection, we will comment on the potential effects of time variations in noise amplitudes.

\begin{figure}[htbp] 
\centering
\includegraphics[clip,width=0.95\columnwidth]{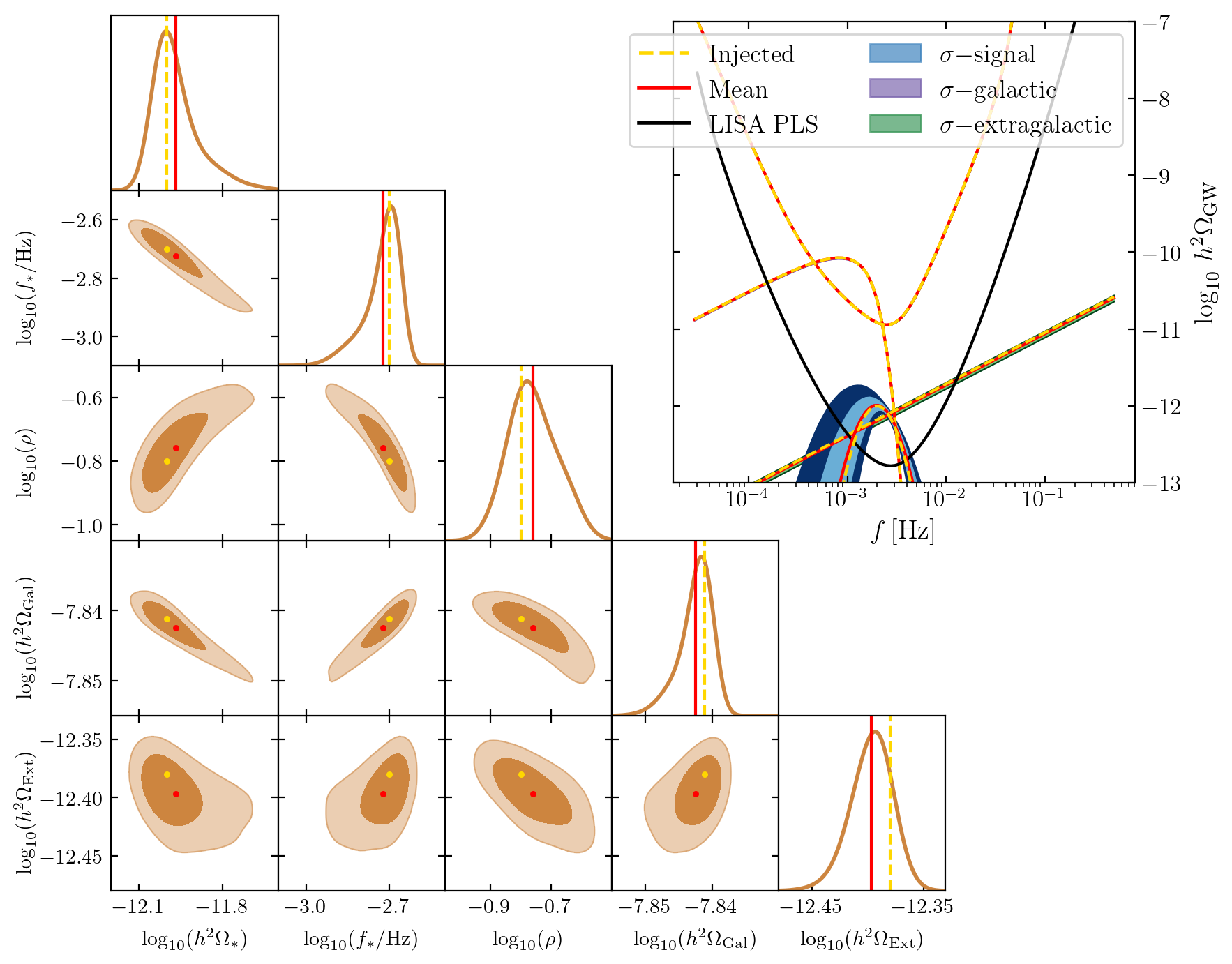}\\
\includegraphics[clip,width=0.5\columnwidth]{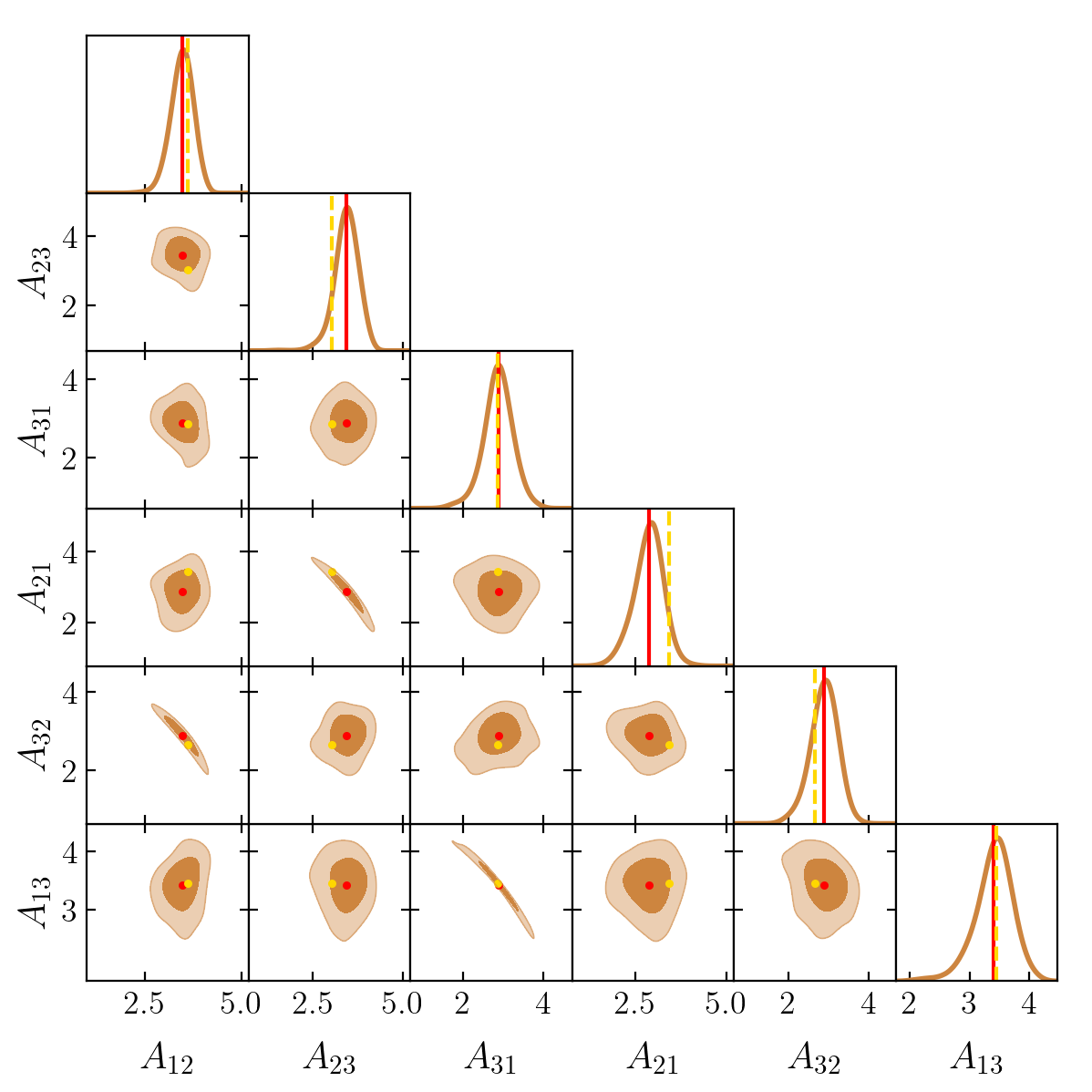}~
\includegraphics[clip,width=0.5\columnwidth]{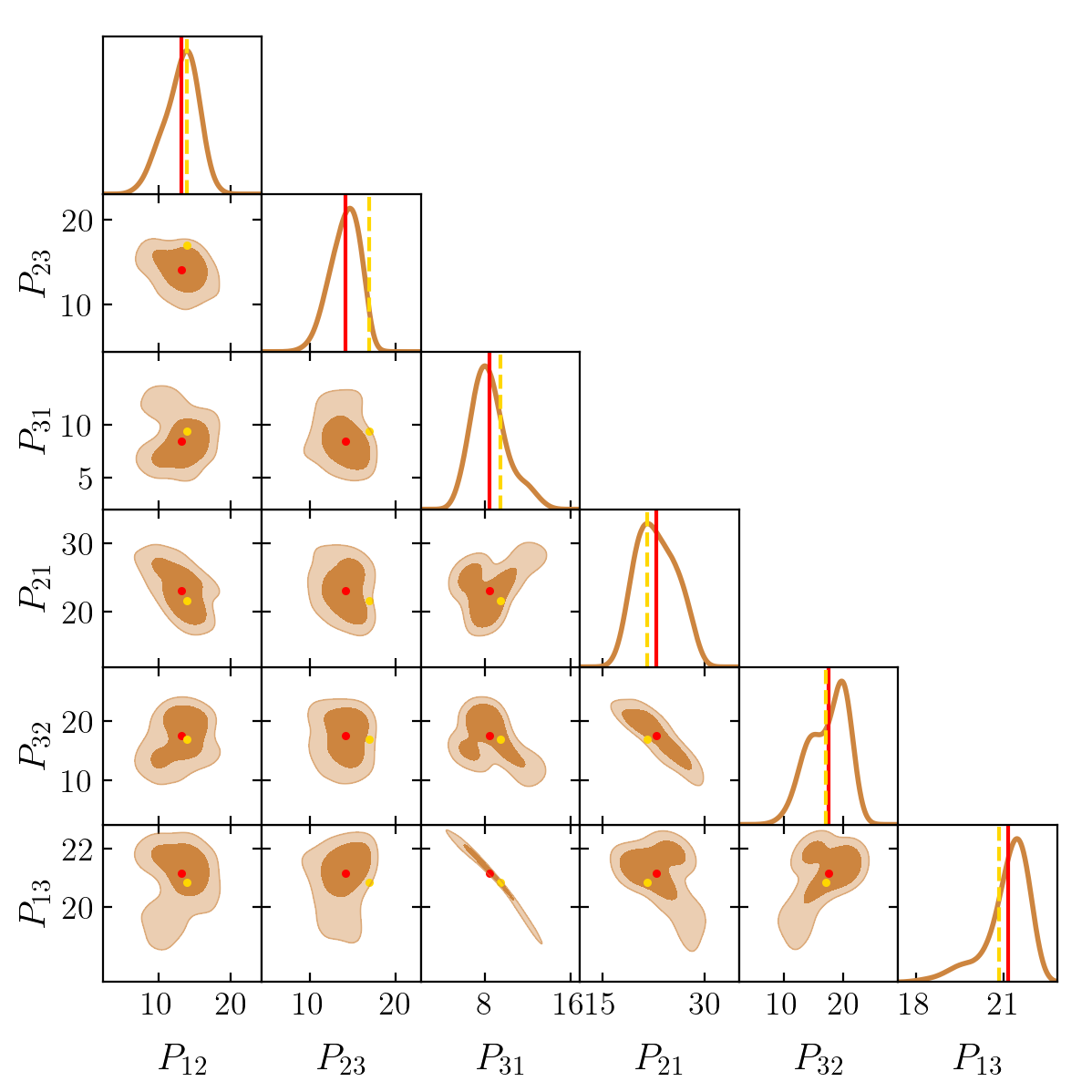}
\caption{
  2D posterior distribution for signal parameters and foreground amplitudes ({\it top panel}), unequal amplitudes of TM ({\it bottom left panel}) and OMS ({\it bottom right panel}), respectively, derived from the data with unequal noise amplitudes. In the corner plot, the yellow and red dots, and corresponding vertical lines, show the injected parameters and their reconstructed mean values. Dark and light orange regions represent  68\% C.L. and 95\% C.L. respectively.
  The top-right inset visualises the injected and reconstructed signals, with 68 and 95\% C.L. error bands, and the LISA PLS (solid black). The error bands on the galactic foreground and instrumental noise are too small to be visible.
  }
    \label{fig:bump_uneq_uneq}
\end{figure}

For the log-normal bump signal, we set the values of parameters as 
\begin{equation}
\{\log_{10}h^2\Omega_*, \log_{10}f_*,\log_{10}\rho_*\} = \{-12,-2.7,-0.8\}
\end{equation}
to have a relatively low signal-to-noise ratio (SNR) $\sim 20$. 
Then, using only the diagonal part of the TDI covariance, we run the analysis for signal parameters, foreground amplitudes, and 12 noise amplitudes.
In Fig.~\ref{fig:bump_uneq_uneq}, we show the triangle plot of the 2D-marginalized posterior of signal parameters,  foreground amplitudes, and unequal noise amplitudes respectively. Since we found no significant correlation between the noise amplitudes and the others, and for aesthetic reasons, we plotted them separately.
For the noise amplitudes, one can see the appearance of degeneracy in $(A_{13}, A_{31}), (A_{21}, A_{23}), (A_{12}, A_{32})$ and $(P_{13}, P_{31})$. 
In fact, this is predicted as they appear similarly to each other in the expressions of the unequal noise spectrum~\eqref{eq:uneq_AET}. We expect these degeneracies to be (at least partially) resolved once the off-diagonal terms are included in the analysis.
In the top-right inset, on the other hand, we visualize the injected and reconstructed signal, noise, and foregrounds with their 68 and 95\% C.L. error bands. As a reference, the power-law sensitivity (PLS)~\cite{Thrane:2013oya, Caprini:2019pxz} for SNR = 10 is shown with the black solid line. Notice that although the reconstruction errors for each noise amplitude parameter are relatively large, the sensitivity itself is precisely determined and error bands are too small to be visible. The errors for the galactic component are also barely visible.

\begin{figure}[htbp] 
\centering
\includegraphics[clip,width=\columnwidth]{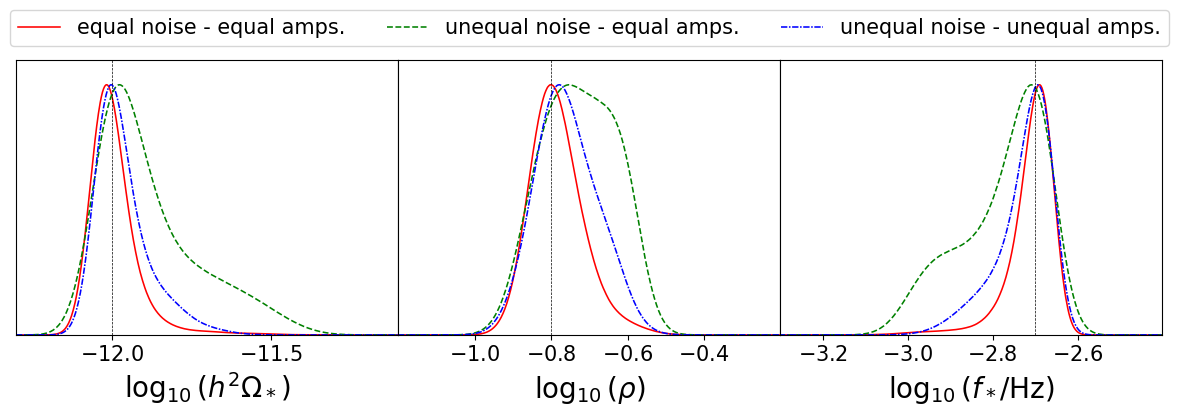}\\
\includegraphics[clip,width=0.9\columnwidth]{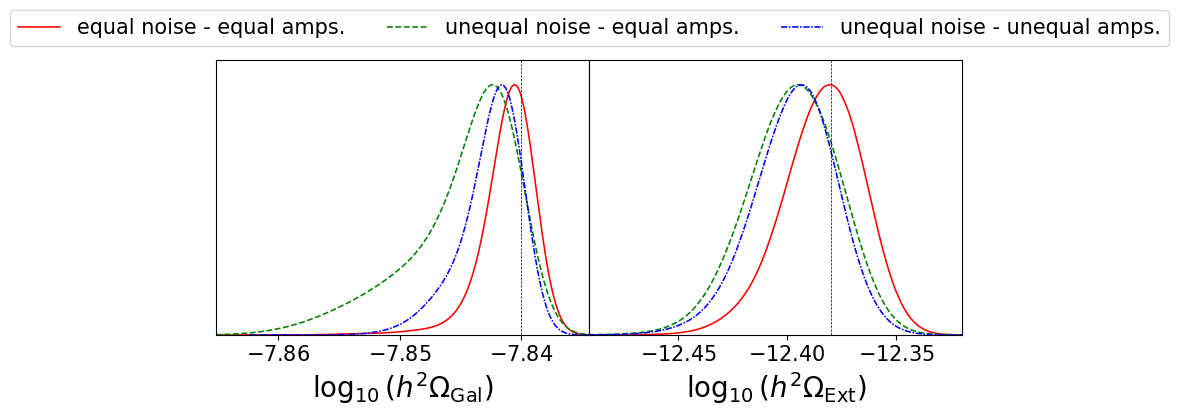}
  \caption{
  1D-marginalized posterior for the signal parameters and foreground amplitudes. Here we compare the three cases: I) equal noise data fitted with equal amplitudes (solid red line), II) unequal noise data fitted with equal amplitudes (green dashed line), III) unequal noise data fitted with unequal amplitudes (blue dot-dashed line). 
  A comparison of I) and II) shows the effect of suboptimality, and a comparison of II) and III) shows the effect of incorrect fitting.
  }
    \label{fig:uneq_1D}
\end{figure}

One key question is to what extent the suboptimality affects the signal parameter estimate. We explored this by generating and analyzing ``idealized'' equal noise data using the same signal and foreground parameter values.
To set a similar SNR with that of unequal noise data, here we set the root mean squared (RMS) value of Eq.~\eqref{eq:unequal_noise_levels} for the equal amplitudes as $(A,P) = (3.19,17.13)$.
As a reference, we also performed equal noise fitting on the unequal noise data.
To compare the uncertainties in the signal parameter reconstruction, we marginalized the posterior over the remaining parameters.
In Fig.~\ref{fig:uneq_1D}, we show the 1D-marginalized posterior of the signal parameters and also foreground amplitudes for the equal noise data fitted with equal amplitudes (red, eq-eq), the unequal noise data fitted with equal amplitudes (green, uneq-eq) and the unequal noise data fitted with unequal amplitudes (blue, uneq-uneq). 
Overall, the posterior distributions are well peaked around the fiducial values.

By comparing the eq-eq (red) and uneq-uneq (blue) cases, we find that the error, or the width of distribution, of signal parameters increases by only a few 10\% at most. 
While there is a similar increase in the uncertainty for $\Omega_{\rm Gal}$, the error for $\Omega_{\rm Ext}$ remains effectively unchanged. 
Although not shown here, a similar behavior was observed for larger values of the signal amplitude, {\it e.g.}, SNR $\simeq 100$.
This result is consistent with Ref.~\cite{Hartwig:2023pft}, which shows that unequal noise amplitudes have little effect on signal reconstruction.
We believe that this is because the signal and noise transfer functions have different structures and frequency dependencies. This was shown, e.g., in Fig.~11 and~13 of Ref.~\cite{Hartwig:2021mzw}, using a flat signal as a proxy. In practice, given the different propagation in the data streams, the signal parameters have very little degeneracy with the noise parameters and complications in the noise model do not affect the measurement of the signal parameters significantly.
Indeed, in the full triangle plot, we did not observe any significant correlation between the noise amplitudes and the signal parameters, or between the noise amplitudes and the foreground amplitudes.
In other words, the perfect knowledge of the noise spectrum makes the signal parameter reconstruction robust against the amplitude inequality.
As we will see below, in Sec.~\ref{sec:extend_fg}, relaxing the assumption on the knowledge of the foreground spectral shape affects more the parameter reconstruction.
We also expect that the reconstruction in the unequal noise and unequal amplitudes case could be improved (and hopefully get closer to the equal noise and equal amplitudes case) once we include the off-diagonal parts of the covariance matrix.

Interestingly, the incorrect fitting, {\it i.e.}, uneq-eq case (green), shows a slightly worse estimate than the uneq-uneq case as the distributions of the former are still concentrated around the injected values and spread out only by a factor of a few.
We believe this is due to the unequal amplitudes~\eqref{eq:unequal_noise_levels} only moderately fluctuating around their RMS values.
With this moderate inequality, AA and EE noise spectra~\eqref{eq:uneq_AET} are well approximated by the equal noise spectrum~\eqref{eq:psdAA} with the RMS amplitudes, which are well determined by the TT-channel.
This situation could be altered if there is a drastic inequality in the amplitude parameters that change by orders of magnitude.
This would make the unequal noise spectrum significantly different from the equal noise spectrum, resulting in a worse and possibly biased estimate in the uneq-eq case.

Finally, let us comment on the time variation of noise amplitudes. After cutting the full data stream into segments sufficiently short for the noise to be effectively stationary, noise non-stationarities would reduce to a modulation of the total noise amplitude on a segment-by-segment basis. As demonstrated in Ref.~\cite{Alvey:2024uoc}, an appropriate strategy that leverages these noise variations would effectively improve the overall constraint power. In this sense, our forecasts on the signal reconstruction errors in Fig.~\ref{fig:uneq_1D} are conservative.

\subsection{Foregrounds with unfixed shape parameters}\label{sec:extend_fg}
Most previous studies included only the foreground amplitudes in the data analysis. For this reason, in this section, we inspect the consequences of going beyond such an assumption and vary all foreground parameters together with the cosmological ones. 
To highlight the effect of unfixed shape parameters, we assumed the equal noise amplitudes $A_{\alpha\beta} = A = 3$ and $P_{\alpha\beta} = P = 15$.

\begin{figure}[htbp] 
\centering
\includegraphics[clip,width=\columnwidth]{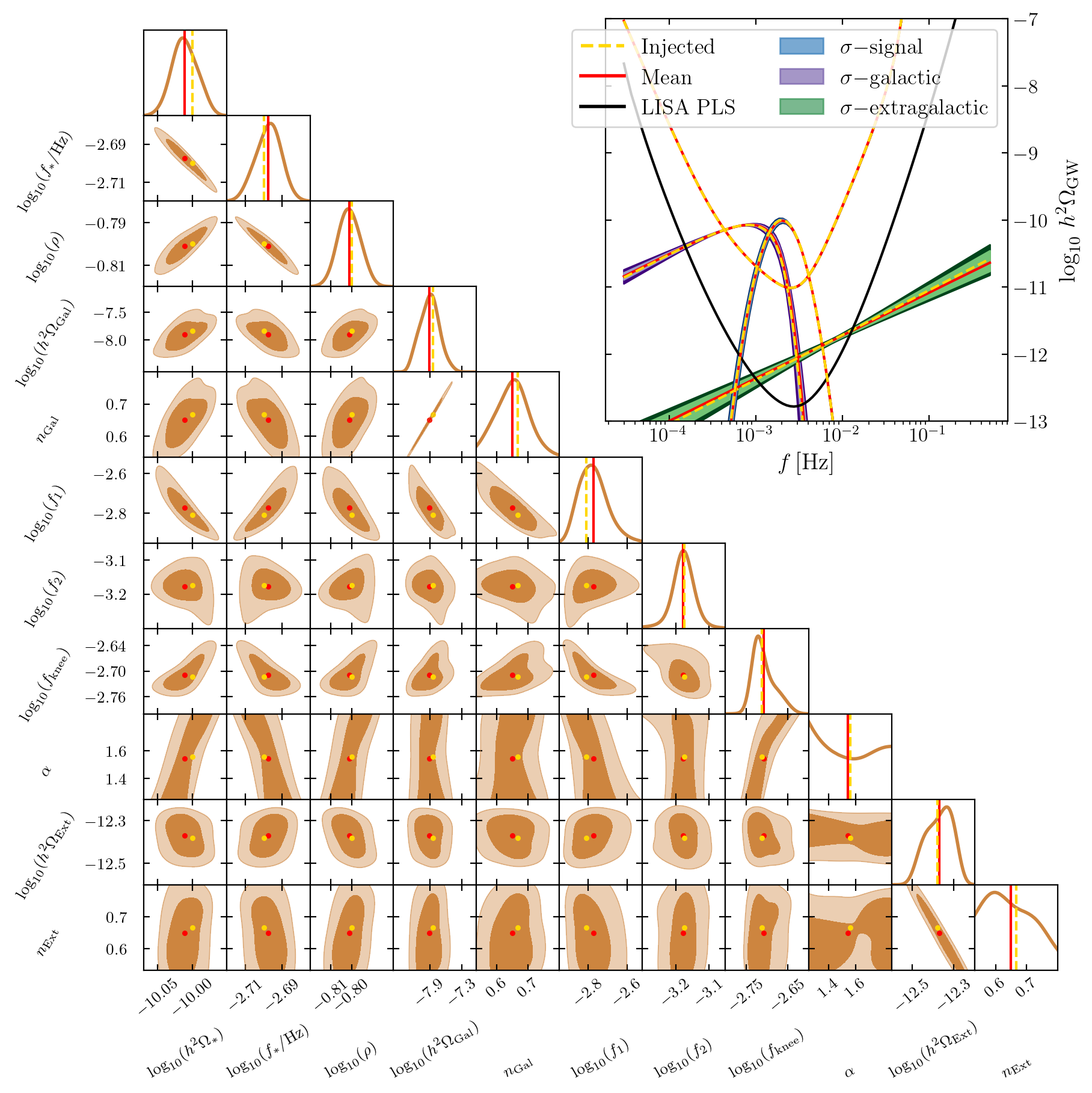}
  \caption{Triangle plot for the signal and foreground parameters including the 6 unfixed shape parameters. The range of log-uniform prior for those parameters is taken to be 20\% of the fiducial values. The color scheme is the same as in Fig.~\ref{fig:bump_uneq_uneq}.}
    \label{fig:bump_fg_ext}
\end{figure}
\begin{figure}[htbp] 
\centering
\includegraphics[clip,width=\columnwidth]{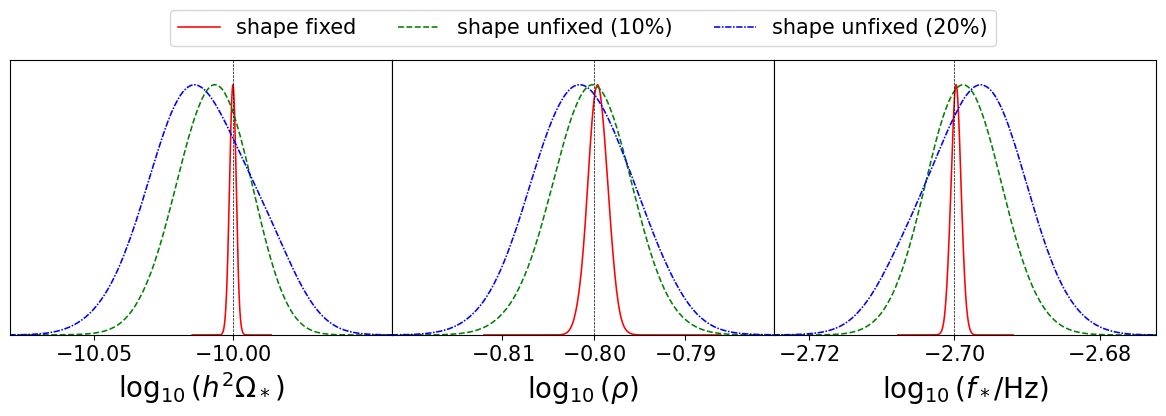}\\
\includegraphics[clip,width=0.67\columnwidth]{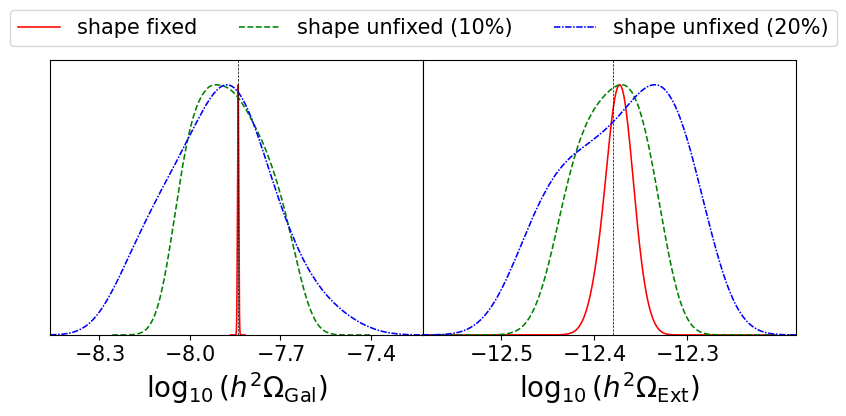}
\vspace{-3mm}
  \caption{1D-marginalized posterior for the signal parameters (top panel) and foreground amplitudes (bottom panel) with foreground shape parameters fixed (red solid line) and unfixed (green dashed line, blue dot-dashed line). 
  The green lines are for the case where 10\% width is assumed while the blue lines are for a 20\% width.}
    \label{fig:extfg_sig_1D}
\end{figure}

For illustrative purposes, we again consider the log-normal bump model as the cosmological signal.
We first analyze the data with signal parameters 
\begin{equation}
\{\log_{10}h^2\Omega_*, \log_{10} f_*, \log_{10}\rho\} = \{-10,-2.7, -0.8\}. \label{eq:sig_model_fg}
\end{equation}
In Ref.~\cite{Karnesis:2021tsh}, the foreground shape parameters, {\it i.e.}, $\{n_{\rm Gal},\ f_1,\ f_2,\ f_{\rm knee},\ \alpha,\ n_{\rm Ext}\}$, were fitted using the MCMC sampling (see Fig.~2 therein), and 68\% C.L. errors of the posterior distributions were at most a few percent of the fiducial value. 
However, our analysis adopts a more conservative approach by considering a scenario with relatively limited information on these parameters. To this end, we perform two different analyses assuming a uniform prior for variables (thus, log-uniform for log-variables) with a width equal to 10\% and 20\% of the fiducial value.

In Fig.~\ref{fig:bump_deg_fg_ext}, we show the full-posterior distribution for the 20\% case. A strong correlation between $\Omega_{\rm Gal}$ and $n_{\rm Gal}$, and between $\Omega_{\rm Ext}$ and $n_{\rm Ext}$ can be observed. 
The correlation between $\Omega_{\rm Ext}$ and $n_{\rm Ext}$ mainly originates from a non-optimal choice of the pivot frequency around which the power-law is anchored, which enhances the degeneracy between the two parameters.
For the galactic foreground, the correlation between $\Omega_{\rm Gal}$ and $n_{\rm Gal}$ suggests that these parameters are mainly determined by the signal in the lower frequency range, which are not significantly influenced by the higher-frequency parameters ${ \alpha, f_1, f_2, f_{\rm knee} }$, that define the high-frequency cut-off. So, at low frequency, we find a power-law like situation, affected by a similar degeneracy as for the extragalactic contribution.

Similarly to Fig.~\ref{fig:uneq_1D}, in Fig.~\ref{fig:extfg_sig_1D}, we compare 1D-posterior for the signal and foreground parameters marginalized over all the other parameters.
In both figures, the red line denotes the case where the foreground shape is fixed. The unfixed cases are represented by the blue and green lines, which are respectively for 10\% and 20\% prior width.
In contrast to the unequal noise case shown in Fig.~\ref{fig:uneq_1D}, we observe a significant increase (up to one order of magnitude depending on the parameters) in the uncertainties of the reconstruction.

\begin{figure}[htbp] 
\centering
\includegraphics[clip,width=\columnwidth]{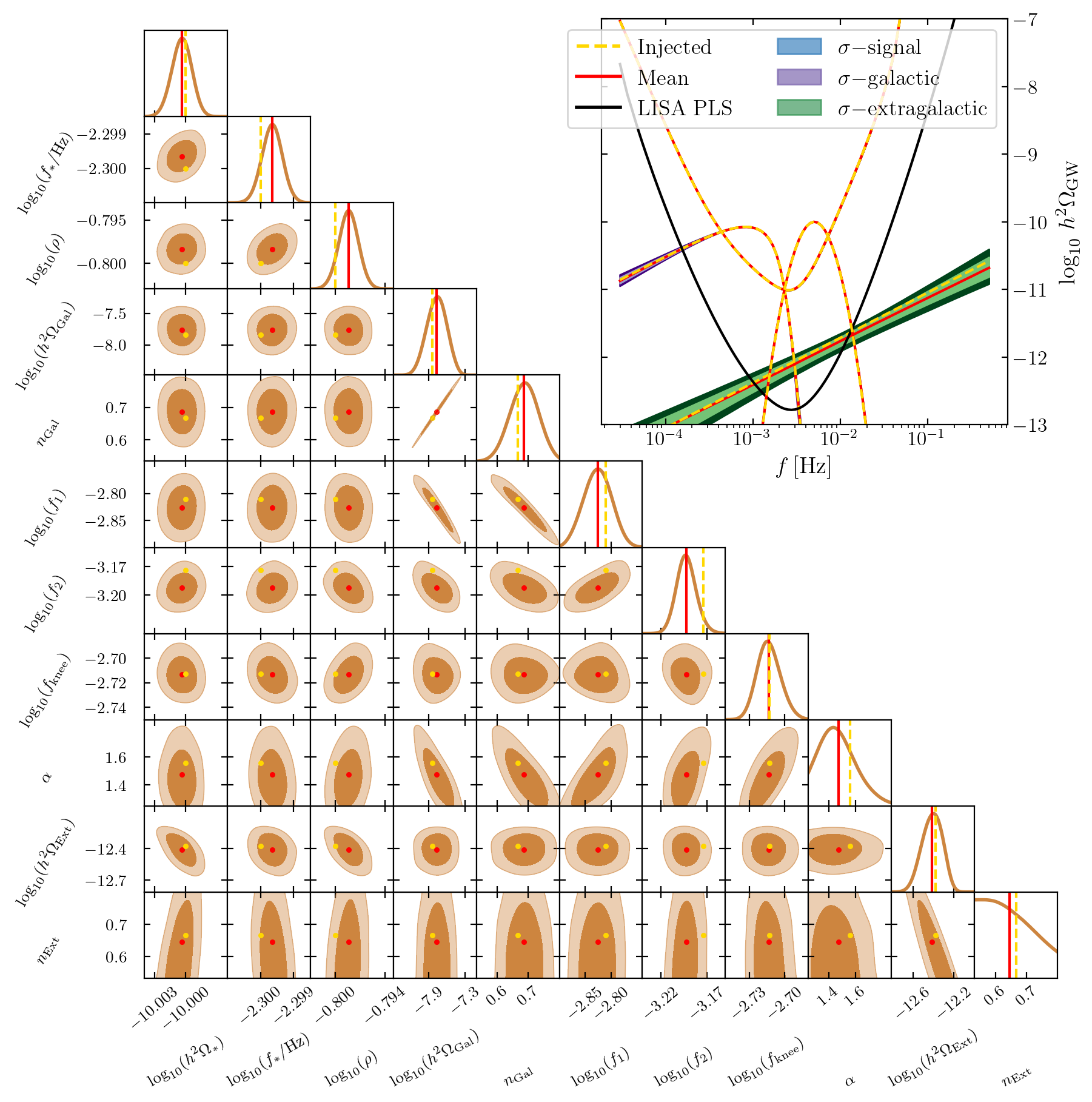}
  \caption{The same plot as Fig.~\ref{fig:bump_fg_ext} but for the signal parameters $\{\log_{10}h^2\Omega_*, \log_{10}f_*,\log_{10}\rho_*\} = \{-10,-2.3,-0.8\}$.}
    \label{fig:bump_nondeg_fg_ext}
\end{figure}

\begin{figure}[htbp] 
\centering
\includegraphics[clip,width=\columnwidth]{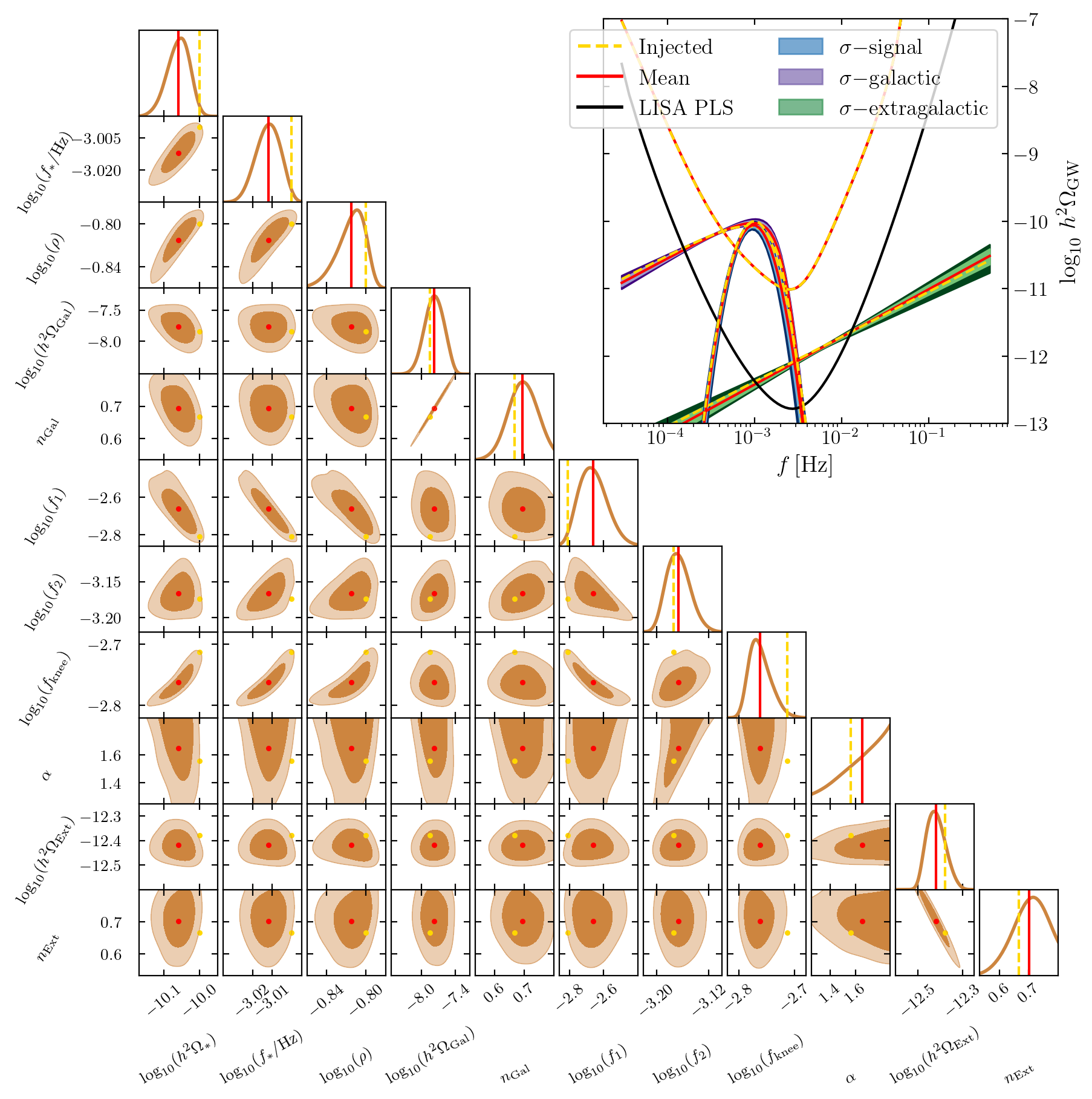}
  \caption{The same plot as Fig.~\ref{fig:bump_fg_ext} but for the signal parameters $\{\log_{10}h^2\Omega_*, \log_{10}f_*,\log_{10}\rho_*\} = \{-10,-3.0,-0.8\}$.}
    \label{fig:bump_deg_fg_ext}
\end{figure}

\begin{figure}[htbp] 
\centering
\includegraphics[clip,width=0.9\columnwidth]{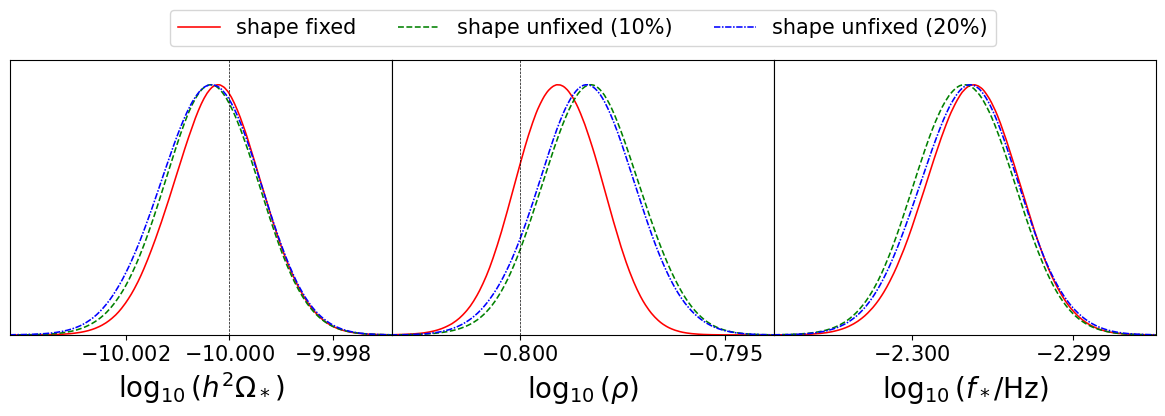}\\
\includegraphics[clip,width=0.6\columnwidth]
{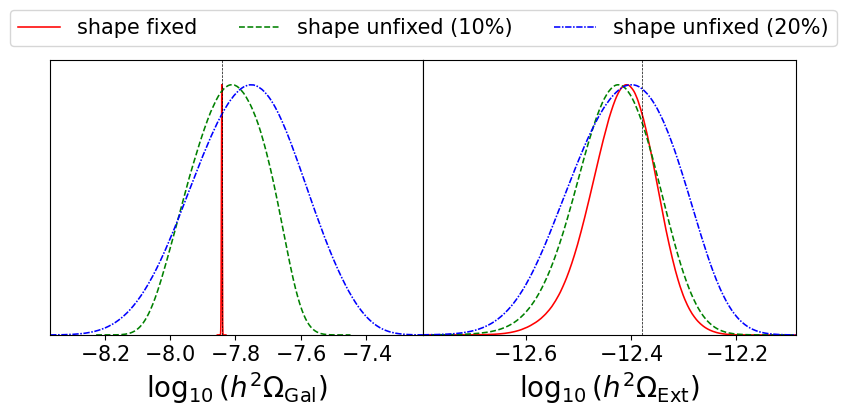}
\vspace{-3mm}
  \caption{1D-marginalized posterior of the signal parameters and foreground amplitudes for the less overlapped case. The color scheme is the same as that in Figs.~\ref{fig:extfg_sig_1D}.}
    \label{fig:extfg_nondeg_1D}
\end{figure}

\begin{figure}[htbp] 
\centering
\includegraphics[clip,width=0.9\columnwidth]{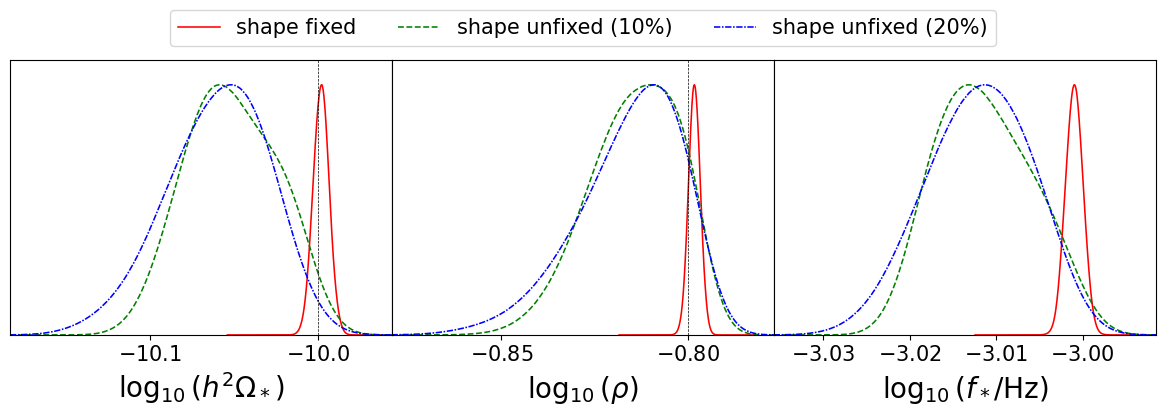}\\
\includegraphics[clip,width=0.6\columnwidth]{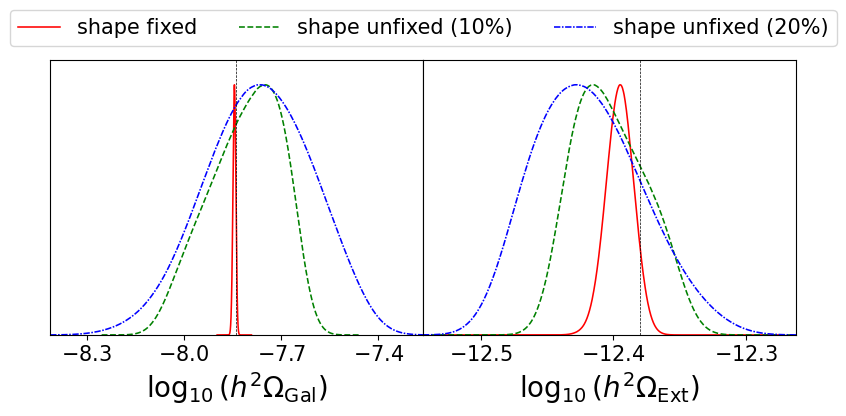}
\vspace{-3mm}
  \caption{The same plot as Fig.~\ref{fig:extfg_nondeg_1D} but for the highly overlapped case.}
    \label{fig:extfg_deg_1D}
\end{figure}

As a complementary to the signal model in Eq.~\eqref{eq:sig_model_fg}, we performed similar analyses for the same log-normal template with varying the peak position $f_*$. In particular, we set $\log_{10}f_* = -2.3, -3.0$ while keeping the other two parameters as $\{\log_{10}h^2\Omega_*, \log_{10}\rho\} = \{-10, -0.8\}$.
We refer to the former choice as the `less overlapped case' and to the latter choice as the `highly overlapped case'. The full triangle plot for the 20\% prior width is shown in Fig.~\ref{fig:bump_nondeg_fg_ext} and Fig.~\ref{fig:bump_deg_fg_ext} (less overlapped and highly overlapped case, respectively).
Similarly to Fig.~\ref{fig:extfg_sig_1D}, we compare marginalized 1D-posterior of signal parameters and foreground amplitudes in Figs.~\ref{fig:extfg_nondeg_1D}--\ref{fig:extfg_deg_1D}. Also, in this case, the increase of error of signal parameters against the prior width is larger.

By comparing Figs.~\ref{fig:extfg_sig_1D},~\ref{fig:extfg_nondeg_1D} and~\ref{fig:extfg_deg_1D}, one can see that the reconstruction error of the signal parameters tends to increase with more overlap between the galactic foreground and the signal.
This is somewhat expected since the high-frequency cut-off of the galactic foreground and the bump signal have quite similar shapes in frequency, making it hard to separate their relative weights.
Indeed, a correspondent increase in the correlation between the signal parameters and the shape parameters $\{ \alpha, \, f_1, \, f_2, \, f_{\rm knee} \}$ is observed in Figs.~\ref{fig:bump_fg_ext},~\ref{fig:bump_nondeg_fg_ext} and~\ref{fig:bump_deg_fg_ext}.

Moreover, we notice that the reconstruction error of the galactic amplitude $\Omega_{\rm Gal}$ is less sensitive to the degree of overlap between the signal and the galactic foreground. 
As mentioned earlier, $\Omega_{\rm Gal}$ is strongly correlated with $n_{\rm Gal}$. 
Since the signals we consider do not mask the lower frequency part, the ability to determine the degenerated set $\{\Omega_{\rm Gal}, \, n_{\rm Gal}\}$ is considered to be comparable in all the three cases.
Therefore, the degradation in the estimation of $\Omega_{\rm Gal}$ is mostly contributed by the degeneracy with $n_{\rm Gal}$ (the same applies to $\Omega_{\rm Ext}$ and $n_{\rm Ext}$). 
We expect that if we consider a broader peak for the signal, covering both the lower and higher frequency parts, we will see a further increase in the reconstruction errors of both the signal parameters and the foreground amplitudes. 
Indeed, in the case of Fig.~\ref{fig:extfg_nondeg_1D} where the bump signal masks the extragalactic component at higher frequencies, the error in $\Omega_{\rm Ext}$ is the largest compared to the other cases where the error is determined solely by the contribution from $n_{\rm Ext}$.
While this discussion generally applies to peaked signals, a dedicated analysis will be needed for signals with more complex spectral shapes.

\section{Discussion}\label{sec:discussion}
In this work, we have studied the effects of unequal noise amplitudes and unfixed foreground shapes on the reconstruction of cosmological SGWBs with LISA.
In previous studies utilizing the \texttt{SGWBinner} code, the following assumptions were made to simplify the analysis: i) static and equal arm lengths, ii) uniform noise amplitudes at each link, and iii) perfect knowledge of the spectral shapes of foregrounds.
Given that the assumptions of equal noise amplitudes and perfect knowledge of spectral shapes are not realistic, we relax assumptions ii) and iii) to test their impact on the capabilities of LISA to measure SGWBs of cosmological origin. 
This more complicated scenario, and, in particular, the higher dimensionality of its parameter space, leads to an increase in the computational cost of the analysis. 
To overcome this, we have updated the \texttt{SGWBinner} code using the \texttt{JAX} library, resulting in the acceleration of the sampling process as summarized in App.~\ref{sec:accelerate}.

While the impact of (static) unequal arm lengths and unequal noise amplitudes was explored in Ref.~\cite{Hartwig:2023pft}, the analysis was limited to a flat power-law signal with relatively large SNR (order 100). For this reason, in the present work, we have extended this analysis using some of the templates for cosmological SGWBs implemented in the \texttt{SGWBinner} code~\cite{Caprini:2024hue,Blanco-Pillado:2024aca,Braglia:2024kpo}. In particular, in Sec.~\ref{sec:uneq_noise}, we have studied the case of a peaked signal with a relatively low SNR (order 10). Consistently with Ref.~\cite{Hartwig:2023pft}, our results show that a moderate (and reasonable) level of inequality between the noise amplitudes does not affect the signal reconstruction significantly. Thus, we conclude that the results of Ref.~\cite{Hartwig:2023pft} hold (with only minor modifications) also for different and more complicated signal shapes, which might also be more feeble. Once again, we stress that, as in Ref.~\cite{Hartwig:2023pft}, our analysis relies on the diagonal terms of the correlation matrix, which makes it suboptimal. We leave the implementation of the full analysis, which requires some code restyling, to future works. 

In Sec.~\ref{sec:extend_fg}, we discussed the impact of less restrictive foreground templates compared to the ones used in previous studies. Again, as a proxy for a cosmological SGWB, we considered a log-normal bump with relatively low SNR, positioned to partially overlap with the high-frequency cut-off of the galactic component. 
In this scenario, we found an order of magnitude increase in the uncertainties on both the signal and foreground parameters compared to the simplified case. The degradation in signal parameter estimations originates from the degeneracy with the parameters controlling the high-frequency cutoff in the galactic component. Meanwhile, the degradation in the determination of the amplitude of the galactic component originates from its fairly strong correlation with the tilt parameter, which primarily arises in the unmasked low frequency part.
Additionally, we observed a reasonable variation in the correlation between the signal and foreground parameters when shifting the peak position of the signal. Specifically, as the signal moves away from the galactic component, it instead begins to mask the extra-galactic component at higher frequencies. As a result, the error in the amplitude of the extra-galactic component, which was otherwise determined by its strong correlation with the tilt parameter, increases due to its correlation with the signal parameters. 
We therefore expect that if the signal peak were broader, covering both lower and higher frequencies, the reconstruction errors in both the signal parameters and the foreground amplitudes would increase further.
Given the significance of the degradation in the parameter uncertainties, these results highlight the importance of precisely modeling the spectrum of the foreground contribution. We stress that while we only run tests with the lognormal signal, we have implemented \texttt{JAX}ed versions of the other templates as well (see~\cref{sec:templates}). We expect the main conclusions drawn from this work to generalize to those templates as well.

Before concluding, we discuss the recent upgrades implemented in the \texttt{SGWBinner} code and outline potential future developments. 
The enhancement of parts of \texttt{SGWBinner} code with the \texttt{JAX} framework allows for GPU/TPU acceleration with the \texttt{XLA} compiler, which can substantially speed up the analysis and facilitates huge data handling. For example, this would be relevant to investigate, {\it e.g.}, anisotropic SGWBs or non-stationarities in the signal or the noise. For previous studies on these topics see~\cite{Contaldi:2020rht,Seoane:2021kkk,LISACosmologyWorkingGroup:2022kbp, Mentasti:2023uyi, Schulze:2023ich, Heisenberg:2024var,Alvey:2024uoc}.
Moreover, \texttt{JAX} compatibility allows for further code upgrades. For example, automatic differentiation, which was not used in the present analysis, provides access to the derivatives of likelihood with respect to any of the parameters. While this could also be useful in the Fisher analysis, it is worth noting that powerful Bayesian inference algorithms such as Hamiltonian MC and variational inference can be used because both methods rely on efficient gradient computation.

\section*{Acknowledgments}
J.K is supported by the JSPS Overseas Research Fellowships.
J.K and M.Pe acknowledge support from Istituto Nazionale di Fisica Nucleare (INFN) through the Theoretical Astroparticle Physics (TAsP) project. M.Pe acknowledges support from the MIUR Progetti di Ricerca di Rilevante Interesse Nazionale (PRIN) Bando 2022 - grant 20228RMX4A, funded by the European Union - Next generation EU, Mission 4, Component 1, CUP C53D23000940006. A.R acknowledges financial support from project BIGA funded by the MUR Progetti di Ricerca di Rilevante Interesse Nazionale (PRIN) Bando 2022 - grant 20228TLHPE - CUP I53D23000630006.  M.Pi acknowledges the hospitality of Imperial College London, which
provided office space during some parts of this project.

\appendix
\section{Noise power spectrum for equal arm length}
\label{sec:full_noise}
In this appendix, we report the expressions of the noise auto- and cross-spectra both in XYZ and AET basis for equal arm length. To simplify the expressions, we define $x \equiv f/f_c$. For the equal noise amplitudes, the total power spectral density for the noise PSDs become~\cite{Flauger:2020qyi,Hartwig:2021mzw,QuangNam:2022gjz}
\begin{equation}
	P_{N,ii}(f, A, P) = 16 \sin^2 x \left\{  \left[3 +\cos (2 x )  \right] S^\text{TM}(f,A) + S^\text{OMS}(f,P) \right\} \; ,
	\label{eq:N_XX}
\end{equation}
and the noise CSDs are
\begin{equation}
	P_{N,ij}(f, A, P) = -8 \sin^2 x \cos x  \left[4 S^\text{TM}(f,A) +  S^\text{OMS}(f,P) \right] \; ,
	\label{eq:N_XY}
\end{equation}
where $i,j\in\{$X,Y,Z$\}$ and $i\ne j$. Therefore, each TDI channel observes equivalent correlated noise in the XYZ basis.
On the other hand, the PSDs in the AET basis read 
\begin{equation}
\begin{aligned}
    P_{N,{\rm AA}}(f, A, P)  & = P_{N,{\rm EE}}(f, A, P) = P_{N,{\rm XX}}(f, A, P) - P_{N,{\rm XY}}(f, A, P) \\
& =  8 \sin^2 x \left\{ 4 \left[1 +\cos x + \cos^2 x\right] S^\text{TM}(f,A) \;  +  \left[ 2 +\cos x \right]S^\text{OMS}(f,P)  \right\}  \; , \\
\end{aligned}\label{eq:psdAA}
\end{equation}
and
\begin{equation}
\begin{aligned}
    P_{N,{\rm TT}}(f, A, P) & = P_{N,{\rm XX}}(f, A, P) + 2 P_{N,{\rm XY}}(f, A, P)\\
    & =  16 \sin^2 x \left\{ 2 \left[ 1 - \cos x \right]^2  S^\text{TM}(f,A) + \left[ 1 - \cos x \right] S^\text{OMS}(f,P) \right\} \; ,
\end{aligned}
\label{eq:psdTT}
\end{equation}
with vanishing CSDs $P_{N,ij}(f, A, P) = 0$ for $i\ne j$ so that the noise covariance matrix is diagonal.

We proceed by reporting the expressions assuming the noise amplitudes at each link might have different values.
In the XYZ basis, the PSDs are given by 
\begin{equation}
\begin{aligned}
P_{N,{\rm XX}}(f) & = 4\sin^2 x \left\{ 4\left[S^{\rm TM}_{21}
+S^{\rm TM}_{31} + (S^{\rm TM}_{12}
+S^{\rm TM}_{13})\cos^2 x \right] + S^{\rm OMS}_{(21)} + S^{\rm OMS}_{(31)} \right\}  \; , \\
P_{N,{\rm YY}}(f)  & = 4\sin^2 x \left\{ 4\left[S^{\rm TM}_{12}
+S^{\rm TM}_{32} + (S^{\rm TM}_{21}
+S^{\rm TM}_{23})\cos^2 x \right] + S^{\rm OMS}_{(12)} + S^{\rm OMS}_{(32)} \right\}  \; , \\
P_{N,{\rm ZZ}}(f)  &= 4\sin^2 x \left\{ 4\left[S^{\rm TM}_{13}
+S^{\rm TM}_{23} + (S^{\rm TM}_{31}
+S^{\rm TM}_{32})\cos^2 x \right] + S^{\rm OMS}_{(13)} + S^{\rm OMS}_{(23)} \right\}  \; ,  
\end{aligned}
\end{equation}
and the CSDs are
\begin{equation}
\begin{aligned}
P_{N,{\rm XY}}(f)  & = - 4\sin^2 x \left[(S^{\rm OMS}_{(12)} + 4S^{\rm TM}_{(12)})\cos x  +iS^{\rm OMS}_{[12]}
\sin x \right]  \; , \\
P_{N,{\rm YZ}}(f)  & = - 4\sin^2 x \left[(S^{\rm OMS}_{(23)} + 4S^{\rm TM}_{(23)})\cos x  +iS^{\rm OMS}_{[23]}
\sin x \right]  \; ,\\
P_{N,{\rm ZX}}(f)  & = - 4\sin^2 x \left[(S^{\rm OMS}_{(31)} + 4S^{\rm TM}_{(31)})\cos x  +iS^{\rm OMS}_{[31]}
\sin x \right]  \; , 
\end{aligned}\label{eq:XYZ_uneq}
\end{equation}
where we define symmetric sum $S^{\rm TM/OMS}_{(\alpha\beta)} \equiv S^{\rm TM/OMS}_{\alpha\beta}+S^{\rm TM/OMS}_{\beta\alpha}$
and anti-symmetric sum $S^{\rm OMS}_{[ij]} \equiv S^{\rm OMS}_{ij} - S^{\rm OMS}_{ji}$.
One can easily check that for the equal level $S^{\rm TM}_{\alpha\beta}(f) = S^{\rm TM}(f,A)$ and $S^{\rm OMS}_{\alpha\beta}(f) = S^{\rm OMS}(f,P)$, Eqs.~\eqref{eq:N_XX} and~\eqref{eq:N_XY} are reproduced.
\texttt{SGWBinner} code generates noise by diagonalizing the noise correlation matrix in XYZ basis. Therefore, all these equations are used to generate the LISA data stream with unequal noise levels.

In the AET basis, the PSDs read
\begin{equation}
\begin{aligned}
P_{N,{\rm AA}}(f)  & = 2\sin^2 x \left\{ 4 \left[(S^{\rm TM}_{21}
+S^{\rm TM}_{23} + S^{\rm TM}_{(31)}) +2S^{\rm TM}_{(31)}\cos x  + (S^{\rm TM}_{12}
+S^{\rm TM}_{32}+S^{\rm TM}_{(31)}) \cos^2 x \right] \;  \right. \\ 
& \hspace{5mm} \left. + \left[(S^{\rm OMS}_{(12)}+S^{\rm OMS}_{(23)}+2S^{\rm OMS}_{(31)}) + 2S^{\rm OMS}_{(31)}\cos x  \right]  \right\} \; , \\
P_{N,{\rm EE}}(f)  & = \frac{2}{3} \sin^2 x \left\{ 4\left[(4S^{\rm TM}_{12} + S^{\rm TM}_{21}
+S^{\rm TM}_{23} + 4S^{\rm TM}_{32} + S^{\rm TM}_{(31)})   \;  \right. \;  \right.\\
& \hspace{5mm} \left.  \left.  
+2(2S^{\rm TM}_{(12)}+2S^{\rm TM}_{(23)}-S^{\rm TM}_{(31)} )\cos x
+ (S^{\rm TM}_{12} + 4S^{\rm TM}_{21} + S^{\rm TM}_{32} +  4S^{\rm TM}_{23} +S^{\rm TM}_{(31)})\cos^2 x \right] \;  \right. \\ 
& \hspace{5mm} \left. + 5S^{\rm OMS}_{(12)}+5S^{\rm OMS}_{(23)}+2S^{\rm OMS}_{(31)} + 2(2S^{\rm OMS}_{(12)}+2S^{\rm OMS}_{(23)}-S^{\rm OMS}_{(31)})\cos x  \right\}  \; ,\\
P_{N,{\rm TT}}(f)  & = \frac{8}{3} \sin^2 x \left\{ 2(S^{\rm TM}_{(12)}+S^{\rm TM}_{(23)}+S^{\rm TM}_{(31)}) \left[1 -\cos  x \right]^2  \;  \right. \\ 
    & \hspace{5mm} \left. + (S^{\rm OMS}_{(12)}+S^{\rm OMS}_{(23)}+S^{\rm OMS}_{(31)})\left[ 1 - \cos x  \right] \right\} \; .
\end{aligned}
\label{eq:uneq_AET}
\end{equation}
Once again, we can see that if we take $S^{\rm TM}_{\alpha\beta}(f) = S^{\rm TM}(f,A)$ and $S^{\rm OMS}_{\alpha\beta}(f) = S^{\rm OMS}(f,P)$, the equal level results in Eqs.~\eqref{eq:psdAA}--~\eqref{eq:psdTT} are reproduced.
For completeness, let us also give the noise CSDs in the AET basis
\begin{equation}
\begin{aligned}
P_{N,{\rm AE}}(f)  & = \frac{2}{\sqrt{3}}\sin^2 x \left[-S^{\rm OMS}_{(12)} + S^{\rm OMS}_{(23)} -4S^{\rm TM}_{21} + 4S^{\rm TM}_{23} - 4S^{\rm TM}_{[31]}\right.\\
& \hspace{5mm}
+2(-S^{\rm OMS}_{(12)} + S^{\rm OMS}_{(23)} - 4S^{\rm TM}_{(12)} + 4S^{\rm TM}_{(23)})\cos x 
+4(-S^{\rm TM}_{12}
+S^{\rm TM}_{32}+S^{\rm TM}_{[31]})\cos^2 x \\
&\left. \hspace{5mm}
-2i\lmk S^{\rm OMS}_{[12]} + S^{\rm OMS}_{[23]} + S^{\rm OMS}_{[31]}\rmk
\sin x \right]  \; , \\
P_{N,{\rm ET}}(f)  & = \frac{2\sqrt{2}}{3}\sin^2 x
\left\{\left[-S^{\rm OMS}_{(12)} - S^{\rm OMS}_{(23)} + 2 S^{\rm OMS}_{(31)} +4 \lmk -2S^{\rm TM}_{12} + S^{\rm TM}_{21} + S^{\rm TM}_{23} -2 S^{\rm TM}_{32} + S^{\rm TM}_{(31)}\rmk \right.\right.\\
&\left. \hspace{5mm}
-4\lmk S^{\rm TM}_{12}-2S^{\rm TM}_{21} -2 S^{\rm TM}_{23} + S^{\rm TM}_{32} + S^{\rm TM}_{(31)}\rmk\cos x  \right] 2\sin^2 \left( \frac{x}{2} \right) \\
&\left. \hspace{5mm} - 3i\lmk S^{\rm OMS}_{[12]} - S^{\rm OMS}_{[23]}\rmk
\sin x \right\}  \;,\\
P_{N,{\rm TA}}(f)  & = \frac{2\sqrt{2}}{\sqrt{3}}\sin^2 x  \left\{\left[-S^{\rm OMS}_{(12)} + S^{\rm OMS}_{(23)} -4\lmk S^{\rm TM}_{21} - S^{\rm TM}_{23} + S^{\rm TM}_{[31]}\rmk \right.\right.\\
&\left. \hspace{5mm}
+4\lmk S^{\rm TM}_{12} - S^{\rm TM}_{32} - S^{\rm TM}_{[31]}\rmk\cos x  \right] 2\sin^2 \left(\frac{x}{2}\right)\\
&\left. \hspace{5mm}
- i\lmk S^{\rm OMS}_{[12]} + S^{\rm OMS}_{[23]} - 2 S^{\rm OMS}_{[31]}\rmk
\sin x \right\} \;,
\end{aligned}\label{eq:AET_uneq}
\end{equation}
where all these vanish for equal noise levels.
We can see the appearance of the terms $S^{\rm TM/OMS}_{[31]}$ and the combinations such as $S^{\rm TM}_{12} - S^{\rm TM}_{32}$ and $S^{\rm TM}_{21} - S^{\rm TM}_{23}$. These could break the degeneracy we found in the suboptimal analysis of unequal noise reported in Sec.~\ref{sec:uneq_noise}.
Note that Eqs.~\eqref{eq:XYZ_uneq} and~\eqref{eq:AET_uneq} agree with the covariance used in Ref.~\cite{Hartwig:2023pft} in the equal arm length limit.

\section{Acceleration in the likelihood computation}\label{sec:accelerate}
Here we report the improvement in the speed of likelihood computations using the \texttt{SGWBinner} code.
We used the speed measurement functionality implemented in  \texttt{Cobaya}~\cite{Torrado:2020dgo}.
This evaluates the computational speed for a few sets of the parameters (points) and then discards one result, returning the speed of likelihood computation in the unit of points per sec.
Notice that by discarding one point, this evaluation does not include the time taken at the initial compilation. 
Therefore, the improvement in evaluated speed, summarized in Tab.~\ref{tab:speed}, corresponds to the overall acceleration of the entire sampling process.

\subsection{Cosmological signal templates}\label{sec:templates}
Here we list the cosmological signal templates used to benchmark the likelihood computation speed.
For more detailed descriptions and the analyses, see Ref.~\cite{Braglia:2024kpo}.

\subsubsection*{Log-normal bump}
\begin{align}
 \label{eq:log-normal_bump}
     h^2 \Omega_{\rm GW} \left( f \right) = h^2\Omega_*\exp\lkk-\frac{1}{2\rho^2}\log^2_{10}\lmk\frac{f}{f_*}\rmk\rkk,
\end{align}
where the parameters $\vec{\theta} = \{\Omega_*, f_*, \rho\}$ control the height, position, and width of the bump, respectively.

\subsubsection*{Excited states}
\begin{equation}
\label{eq:templateexcited}
h^2\Omega_{\mathrm{GW}}^{\rm ES}(f, \vec{\theta}_{\rm cosmo}) = 
\frac{h^2 \Omega_*}{0.052} \frac{1}{y^3}\left[1-\frac{y^2}{4\gamma_{\rm ES}^2}\right]^2\left[\sin(y) -4\frac{\sin^2(y/2)}{y}\right]^2  \Theta\!\left(2\gamma_{\rm ES}  - y \right),
\end{equation}
where we introduce $y \equiv f\omega_{\rm ES}/2$. The parameters are $\vec{\theta}_{\rm cosmo} = \{\Omega_*, \gamma_{\rm ES}, \omega_{\rm ES}\}$.

\subsubsection*{Resonant oscillations}
\begin{eqnarray}
    h^2 \Omega_{\textrm{GW}}^{\textrm{RO}}(f,\vec{\theta}_{\rm cosmo}) &=&  \Big\{1+ \mathcal{A}_1(A_{\log},\omega_{\rm log}) \cos \big[\omega_\textrm{log} \ln (f/\textrm{Hz}) + \theta_{\textrm{log},1} \big]
    \label{eq:resonant-template}\\
    &&\hspace{.53cm}+ \mathcal{A}_2(A_{\log},\omega_{\rm log})  \cos \big[2 \omega_\textrm{log} \ln (f/\textrm{Hz}) + \theta_{\textrm{log},2} \big] \Big\} 
    h^2\Omega^{\rm env}_{\textrm{GW}}(f,\vec{\theta}_{\rm env})\,,\nonumber
    \label{eq:resonant-template2}
\end{eqnarray}
with~\cite{Fumagalli:2021cel} %
\begin{align}
    \label{eq:resonant-template-const} 
    \mathcal{A}_{1} &= \frac{A_{\textrm{log}} \mathcal{C}_1(\omega_\textrm{log})}{1 + A_{\textrm{log}}^2 \mathcal{C}_0(\omega_\textrm{log})},\qquad
     \theta_{\textrm{log},1}= \phi_\textrm{log}+ \theta_{\textrm{log},1}(\omega_\textrm{log}),  \\
    \nonumber \mathcal{A}_{2} &= \frac{A_{\textrm{log}}^2 \mathcal{C}_2(\omega_\textrm{log})}{1 + A_{\textrm{log}}^2 \mathcal{C}_0(\omega_\textrm{log})},\qquad
    \theta_{\textrm{log},2}= 2 \phi_\textrm{log} + \theta_{\textrm{log},2}(\omega_\textrm{log}) \, ,
\end{align}
where $\mathcal{C}_{0,1,2}(\omega_\textrm{log})$ and $\theta_{\textrm{log},1,2}(\omega_\textrm{log})$ are numerical functions that depend 
on the cosmic expansion at the time the SGWB was produced. The parameters of this model are $\vec{\theta}_{\rm cosmo} = \{\Omega_*, A_{\log}, \omega_{\log},\phi_{\log}\}$ 

\subsubsection*{Broken power law}
\begin{align}
 \label{eq:Broken_pl}
     h^2 \Omega_{\rm GW}^{\rm BPL} \left( f \right) = h^2\Omega_* \left(\frac{f}{f_{*}}\right)^{{n_t}_1 } \left[\frac{1+\left(f/f_{*}\right)^{1/\delta}}{2}\right]^{\delta ({n_t}_2- {n_t}_1)},
\end{align}
where the parameters are $\vec{\theta}_{\rm cosmo} = \{ \Omega_*, f_*, {n_t}_1,{n_t}_2, \delta \}$.

\subsubsection*{Double peak}
\begin{equation} 
\begin{aligned}
h^2\Omega^{\rm DP}_{\rm GW}(f, \vec{\theta}_{\rm cosmo}) =& 
\,\,h^2 \Omega_*
\Bigg [\beta \, \left( \frac{f}{\kappa_1 f_*} \right)^{\!n_{p}}  \left[\frac{c_1 - f/ f_*}{c_1 - \kappa_1}\right]^{\!\!\frac{n_{p}}{\kappa_1} \left(c_1-\kappa_1\right)} \Theta\!\left(c_1 - \frac{f}{f_*} \right) \\
&+ \exp\!\left[-\frac{1}{2\rho^2} \log_{10}^2\!\left(\frac{f}{\kappa_2 f_*}\right)\right] \left\{1 + {\rm erf} \left[ -\gamma \log_{10}\!\left(\frac{f}{\kappa_2 f_*}\right) \right] \right\}
\Bigg ] 
\end{aligned}
\label{eq:templateDP} \;,
\end{equation}
where it has seven parameters in total as $\vec{\theta}_{\rm cosmo} = \{ \Omega_*, f_*, \beta, \kappa_1, \kappa_2, \rho, \gamma \}$.

\subsection{Likelihood computation speed}
We applied the speed measurement of likelihood computation described above to the templates listed in App.~\ref{sec:templates}.
In Tab.~\ref{tab:speed}, we summarize the measured speed both for the previous version and for our updated code.

\begin{table}[htbp]
\caption{Summary of the measured speed in the likelihood computation. Here $N_{\rm par}$ represents the number of parameters and the speed is in the unit of points per second.
}\label{tab:speed}
\begin{tabular}{ccccc}
 \\[-1ex]
 \multicolumn{5}{c}{\textbf{Measured speed (per second) of likelihood computation}}\\[0.5ex] 
 \hline
 Template & description & $N_{\rm par}$ & old & updated\\ [0.5ex] 
 \hline\hline\\
 Power law
 &
 a simple power law with fixed pivot frequency
 &
 2
 &
 341
 & 3630\\ \\[-1.5ex]
 \hline\\[-1ex]

 Log-normal bump
 &
 bump with log-normal shape (Eq.~\eqref{eq:log-normal_bump})
 &
 3
 &
 334
 & 3830
 \\ \\[-1.5ex]
 \hline\\[-1ex]

 Excited states
 &
 bump with periodic sub-peaks (Eq.~\eqref{eq:templateexcited})
 &
 3
 &
 315
 & 3340
 \\ \\[-1.5ex]
 \hline\\[-1ex]
  
 Resonant oscillations
 &
 log-periodic oscillations (Eq.~\eqref{eq:resonant-template})
 &
 4
 &
 314
 & 2350\\ \\[-1.5ex]
 \hline\\[-1ex]

 Broken power law
 &
 power-law changing its slope (Eq.~\eqref{eq:Broken_pl})
 &
 5
 &
 318
 &
 1790\\ \\[-1.5ex]
 \hline\\[-1ex]
 
 Double peak
 &
 skewed peak at higher frequency (Eq.~\eqref{eq:templateDP})
 &
 7
 &
 316
 &
 1410\\ \\[-1.5ex]
 \hline\\[-1ex]
 
\end{tabular}
\end{table}
One can see that a factor of 10 acceleration is achieved with \texttt{JAX}-ed code for the first three examples.
However, the gain in the broken power law model and the double peak model are not as significant as the others.
On one hand, this may be due to the increased number of parameters. 
On the other hand, we found that the  \texttt{JAX}-ed code cannot accelerate as much in computing powers of arrays.
This is because \texttt{XLA} is a compiler designed to speed up linear algebra. Therefore, even with the \texttt{JAX}-ed code, sampling for templates involving powers of an array in a complex manner takes a relatively longer time. 

Finally, we note that all these computations are performed on the CPU for compressed data. \texttt{JAX} outperforms \texttt{NumPy} especially when the code is run on GPU to work with huge data. In such a situation, {\it e.g.}, when the \texttt{SGWBinner} code takes into account non-stationarity, the \texttt{JAX}-ed code would show better performance also for complex templates including the last two examples.

\bibliographystyle{JHEP}
\bibliography{references}

\end{document}